\documentclass[10pt,a4paper]{article}
\usepackage{amsmath,amssymb,amsfonts}
\usepackage{graphicx}
\usepackage{hyperref}
\usepackage[margin=0.75in]{geometry}
\usepackage{float}
\usepackage{bm}
\usepackage{amsthm}
\usepackage{mathrsfs}
\usepackage{booktabs}
\usepackage{algorithm}
\usepackage{algorithmicx}
\usepackage{algpseudocode}

\usepackage{setspace}
\usepackage{titlesec}
\titlespacing*{\section}{0pt}{8pt}{4pt}
\titlespacing*{\subsection}{0pt}{6pt}{3pt}
\titlespacing*{\subsubsection}{0pt}{5pt}{2pt}

\AtBeginDocument{
  \setlength{\abovedisplayskip}{4pt}
  \setlength{\belowdisplayskip}{4pt}
  \setlength{\abovedisplayshortskip}{2pt}
  \setlength{\belowdisplayshortskip}{2pt}
}

\let\oldbibliography\thebibliography
\renewcommand{\thebibliography}[1]{%
  \oldbibliography{#1}%
  \setlength{\itemsep}{0pt}%
  \setlength{\parskip}{0pt}%
}

\title{\texttt{amangkurat}: A Python Library for Symplectic Pseudo-Spectral Solution of the Idealized (1+1)D Nonlinear Klein-Gordon Equation}

\author{Sandy H. S. Herho$^{1,2,*}$ and Siti N. Kaban$^{3}$}

\date{}

\begin{document}
\maketitle

\begin{center}
\small
$^{1}$Department of Earth and Planetary Sciences, University of California, Riverside, CA 92521, USA\\
$^{2}$School of Systems Science and Industrial Engineering, State University of New York, Binghamton, NY 13902, USA\\
$^{3}$Financial Engineering Program, WorldQuant University, Washington, D.C. 20002, USA\\
$^{*}$e-mail: sandy.herho@email.ucr.edu
\end{center}

\begin{abstract}
\noindent This study introduces \texttt{amangkurat}, an open-source Python library designed for the robust numerical simulation of relativistic scalar field dynamics governed by the nonlinear Klein-Gordon equation in $(1+1)$D spacetime. The software implements a hybrid computational strategy that couples Fourier pseudo-spectral spatial discretization with a symplectic St\o rmer-Verlet temporal integrator, ensuring both exponential spatial convergence for smooth solutions and long-term preservation of Hamiltonian structure. To optimize performance, the solver incorporates adaptive timestepping based on Courant-Friedrichs-Lewy (CFL) stability criteria and utilizes Just-In-Time (JIT) compilation for parallelized force computation. The library's capabilities are validated across four canonical physical regimes: dispersive linear wave propagation, static topological kink preservation in phi-fourth theory, integrable breather dynamics in the sine-Gordon model, and non-integrable kink-antikink collisions. Beyond standard numerical validation, this work establishes a multi-faceted analysis framework employing information-theoretic entropy metrics (Shannon, R\'{e}nyi, and Tsallis), kernel density estimation, and phase space reconstruction to quantify the distinct phenomenological signatures of these regimes. Statistical hypothesis testing confirms that these scenarios represent statistically distinguishable dynamical populations. Benchmarks on standard workstation hardware demonstrate that the implementation achieves high computational efficiency, making it a viable platform for exploratory research and education in nonlinear field theory.
\end{abstract}

\section{Introduction}

The nonlinear Klein-Gordon equation governs relativistic scalar field dynamics in $(1+1)$-dimensional spacetime and has emerged as a fundamental model for understanding topological solitons, breathers, and collision phenomena across remarkably diverse physical contexts~\cite{Perring1962,Dashen1975}. Solitonic solutions---spatially localized, finite-energy field configurations that propagate without dispersion---arise naturally in systems with nonlinear self-interaction potentials, including $\phi^4$ theory (domain walls, kinks) and the sine-Gordon equation (Josephson junctions, magnetic flux lines, structural dislocations)~\cite{Barone1971,Bishop1980}. These structures exhibit remarkable stability conferred by topological charge conservation~\cite{Manton1987,Shnir2018}, rendering them robust against perturbations and relevant to phenomena ranging from cosmological defect formation~\cite{Zurek1985} to condensed matter phase transitions~\cite{Ranada1977}. Beyond equilibrium soliton profiles, collision dynamics between kinks and antikinks reveal fascinating non-integrable behavior, including resonant multi-bounce windows, breather formation, and fractal structure in outcome parameter space~\cite{Campbell1983,Anninos1991,Goodman2005}---phenomena notably absent in integrable systems where inverse scattering methods guarantee elastic collisions~\cite{Scott1973}.

Numerical solution of the Klein-Gordon equation presents interesting computational challenges: achieving spatial accuracy for resolving sharp soliton gradients (characteristic widths $\xi \sim 1$--$10$ in natural units) while maintaining temporal fidelity for preserving Hamiltonian structure over long integration times $t \gg 100$. Traditional finite-difference methods, while straightforward to implement, can exhibit numerical dissipation and dispersion effects that accumulate over time~\cite{Courant1928}. Pseudo-spectral (Fourier) methods offer superior exponential convergence $\mathcal{O}(\exp(-cN_x))$ for smooth solutions, dramatically reducing required grid resolution relative to finite differences~\cite{Fornberg2009,Trefethen2000}, naturally accommodating periodic boundary conditions suitable for many physical scenarios. Temporal discretization via symplectic integrators---schemes that preserve a discrete symplectic structure approximating the continuous Hamiltonian flow---provides excellent bounded energy behavior over extended time periods~\cite{Hairer2003,Yoshida1990}, offering substantial advantages over conventional methods where energy drift can accumulate. The St\o{}rmer-Verlet (leapfrog) scheme, a robust second-order symplectic method, delivers an optimal balance between accuracy and computational efficiency~\cite{Leimkuhler2009}, with natural pathways to higher-order variants for applications requiring enhanced precision.

Despite the considerable physical and mathematical maturity of soliton theory, accessible numerical implementations combining spectral spatial accuracy with symplectic temporal integration have remained relatively scarce in the open-source ecosystem. Existing tools often require commercial software (e.g. COMSOL, Mathematica, MATLAB), impose restrictive licenses, or lack comprehensive documentation for reproducible research. Python has become the leading language for scientific computing due to its extensive numerical libraries (NumPy~\cite{Harris2020}, SciPy~\cite{Virtanen2020}), accessible syntax facilitating rapid prototyping and pedagogical applications, and mature ecosystem for data management (NetCDF4~\cite{Rew1990}), visualization (Matplotlib~\cite{Hunter2007}), and performance optimization (Numba just-in-time (JIT) compilation~\cite{Lam2015}). For $(1+1)$-dimensional problems with grid sizes $N_x \sim 10^2$--$10^3$ and integration times $t \sim 10^1$--$10^2$, Python performance with Numba acceleration proves highly suitable for exploratory research, educational demonstrations, and efficient prototyping prior to production-scale implementation~\cite{Herho2024Eks,Herho2025Schrodinger,Herho2025KH2D}.

This work introduces \texttt{amangkurat}, an open-source Python library implementing a pseudo-spectral St\o{}rmer-Verlet solver for the idealized $(1+1)$-dimensional nonlinear Klein-Gordon equation. The solver effectively combines Fourier-space Laplacian evaluation with symplectic leapfrog time-stepping, adaptive CFL-limited timestep control, and optional Numba parallelization for multi-core execution. The current implementation employs periodic spatial boundary conditions and uniform grids, design choices that streamline the computational framework while accommodating a wide range of physically relevant scenarios. Adaptive timestepping utilizes a practical heuristic criterion based on field magnitude, and the implementation focuses on providing a well-tested proof-of-concept platform that can be readily extended with additional validation features as needed for specific applications. Outputs include self-describing NetCDF4 files with comprehensive metadata and animated GIF visualizations enabling effective qualitative assessment. We demonstrate the solver on four canonical scenarios: dispersive linear wave propagation ($V(\phi) = \frac{1}{2}m^2\phi^2$), static topological kink preservation ($\phi^4$ theory), breather dynamics (sine-Gordon equation), and kink-antikink collisions. Statistical characterization via information-theoretic entropy metrics, kernel density estimation, and phase space reconstruction successfully quantifies spatial coherence, amplitude distributions, and dynamical signatures across these regimes. Complete source code, configuration files, and supplementary data are publicly archived to facilitate reproducibility and extension by the research community, with clear pathways for incorporating additional diagnostic capabilities such as explicit energy conservation monitoring and systematic convergence analysis for specialized research applications.

\section{Methods}

\subsection{Mathematical Formulation}

We begin with the fundamental variational principle governing scalar field dynamics in $(1+1)$-dimensional Minkowski spacetime. Consider a real-valued scalar field $\phi: \mathbb{R}^{1,1} \to \mathbb{R}$ defined on the spacetime manifold equipped with metric signature $\eta_{\mu\nu} = \text{diag}(-1,+1)$ in natural units where $c = \hbar = 1$. The field dynamics emerges from extremizing the action functional $S[\phi]$ constructed from the Lagrangian density $\mathcal{L}(\phi, \partial_\mu\phi)$. For a relativistic scalar field with self-interaction potential $V(\phi)$, the Lagrangian density takes the canonical form~\cite{Goldstein2002,Landau1975}
\begin{equation}
\mathcal{L} = \frac{1}{2}\eta^{\mu\nu}\partial_\mu\phi\partial_\nu\phi - V(\phi),
\label{eq:lagrangian}
\end{equation}
where the kinetic term incorporates both temporal and spatial derivatives with appropriate metric factors. Expanding the metric contraction explicitly with $\eta^{\mu\nu}\partial_\mu\phi\partial_\nu\phi = \eta^{00}\partial_0\phi\partial_0\phi + \eta^{11}\partial_1\phi\partial_1\phi = -(\partial_t\phi)^2 + (\partial_x\phi)^2$, we obtain
\begin{equation}
\mathcal{L} = \frac{1}{2}\left(\frac{\partial \phi}{\partial t}\right)^2 - \frac{1}{2}\left(\frac{\partial \phi}{\partial x}\right)^2 - V(\phi).
\label{eq:lagrangian_expanded}
\end{equation}

The action functional is defined as the spacetime integral
\begin{equation}
S[\phi] = \int_{\mathbb{R}} dt \int_{\mathbb{R}} dx\, \mathcal{L}(\phi, \partial_\mu\phi),
\label{eq:action}
\end{equation}
where we assume sufficient decay conditions $|\phi(x,t)| \to 0$ and $|\partial_\mu\phi(x,t)| \to 0$ as $|x| \to \infty$ to ensure convergence of the spatial integrals. Hamilton's principle of stationary action $\delta S[\phi] = 0$ under arbitrary infinitesimal field variations $\delta\phi(x,t)$ with compact support yields the equations of motion. Computing the variation explicitly, we write
\begin{equation}
\delta S = \int dt \int dx\, \delta\mathcal{L} = \int dt \int dx \left[\frac{\partial\mathcal{L}}{\partial\phi}\delta\phi + \frac{\partial\mathcal{L}}{\partial(\partial_\mu\phi)}\delta(\partial_\mu\phi)\right].
\label{eq:variation_action}
\end{equation}

For the second term, we recognize that the variation commutes with differentiation: $\delta(\partial_\mu\phi) = \partial_\mu(\delta\phi)$. Substituting this relation gives
\begin{equation}
\delta S = \int dt \int dx \left[\frac{\partial\mathcal{L}}{\partial\phi}\delta\phi + \frac{\partial\mathcal{L}}{\partial(\partial_\mu\phi)}\partial_\mu(\delta\phi)\right].
\label{eq:variation_step1}
\end{equation}

Integration by parts on the second term yields
\begin{align}
\int dt \int dx\, \frac{\partial\mathcal{L}}{\partial(\partial_\mu\phi)}\partial_\mu(\delta\phi) &= \int dt \int dx\, \partial_\mu\left[\frac{\partial\mathcal{L}}{\partial(\partial_\mu\phi)}\delta\phi\right] \nonumber \\
&\quad - \int dt \int dx\, \left[\partial_\mu\left(\frac{\partial\mathcal{L}}{\partial(\partial_\mu\phi)}\right)\right]\delta\phi.
\label{eq:integration_by_parts}
\end{align}

The first term on the right-hand side vanishes as a total divergence upon application of the divergence theorem, given our boundary conditions that $\delta\phi$ has compact support and vanishes at spatial and temporal infinities. Thus we obtain
\begin{equation}
\delta S = \int dt \int dx\, \delta\phi \left[\frac{\partial\mathcal{L}}{\partial\phi} - \partial_\mu\left(\frac{\partial\mathcal{L}}{\partial(\partial_\mu\phi)}\right)\right].
\label{eq:variation_step2}
\end{equation}

Since this variation must vanish for arbitrary $\delta\phi(x,t)$, the fundamental lemma of the calculus of variations requires that the integrand vanish identically, yielding the Euler-Lagrange equation
\begin{equation}
\partial_\mu\left(\frac{\partial\mathcal{L}}{\partial(\partial_\mu\phi)}\right) - \frac{\partial\mathcal{L}}{\partial\phi} = 0.
\label{eq:euler_lagrange}
\end{equation}

We now evaluate the requisite partial derivatives from the Lagrangian density. From Eq.~\eqref{eq:lagrangian}, the derivative with respect to $\partial_\mu\phi$ gives
\begin{equation}
\frac{\partial\mathcal{L}}{\partial(\partial_\mu\phi)} = \frac{1}{2}\frac{\partial}{\partial(\partial_\mu\phi)}\left[\eta^{\alpha\beta}\partial_\alpha\phi\partial_\beta\phi\right] = \frac{1}{2}\left[\eta^{\alpha\beta}\delta_\mu^\alpha\partial_\beta\phi + \eta^{\alpha\beta}\partial_\alpha\phi\delta_\mu^\beta\right] = \frac{1}{2}\left[\eta^{\mu\beta}\partial_\beta\phi + \eta^{\alpha\mu}\partial_\alpha\phi\right].
\label{eq:derivative_kinetic}
\end{equation}

Using the symmetry of the metric tensor $\eta^{\mu\nu} = \eta^{\nu\mu}$, both terms are identical, yielding
\begin{equation}
\frac{\partial\mathcal{L}}{\partial(\partial_\mu\phi)} = \eta^{\mu\nu}\partial_\nu\phi.
\label{eq:canonical_momentum_density}
\end{equation}

The derivative with respect to the field itself is simply
\begin{equation}
\frac{\partial\mathcal{L}}{\partial\phi} = -\frac{\partial V}{\partial\phi} = -V'(\phi),
\label{eq:derivative_potential}
\end{equation}
where the prime denotes differentiation with respect to the field argument.

Substituting Eqs.~\eqref{eq:canonical_momentum_density} and \eqref{eq:derivative_potential} into the Euler-Lagrange equation~\eqref{eq:euler_lagrange}, we obtain
\begin{equation}
\partial_\mu\left(\eta^{\mu\nu}\partial_\nu\phi\right) + V'(\phi) = 0.
\label{eq:field_equation_covariant_1}
\end{equation}

Expanding the covariant derivative operator yields
\begin{equation}
\eta^{\mu\nu}\partial_\mu\partial_\nu\phi + V'(\phi) = 0.
\label{eq:field_equation_covariant_2}
\end{equation}

Defining the d'Alembertian operator (or wave operator) as the metric contraction of second derivatives
\begin{equation}
\Box \equiv \eta^{\mu\nu}\partial_\mu\partial_\nu = \eta^{00}\partial_0\partial_0 + \eta^{11}\partial_1\partial_1 = -\partial_t^2 + \partial_x^2,
\label{eq:dalembertian}
\end{equation}
we arrive at the compact covariant form
\begin{equation}
\Box\phi + V'(\phi) = 0.
\label{eq:kg_covariant}
\end{equation}

Expanding explicitly in coordinates $(t,x)$ with our metric signature yields the nonlinear Klein-Gordon equation
\begin{equation}
\frac{\partial^2 \phi}{\partial t^2} - \frac{\partial^2 \phi}{\partial x^2} + V'(\phi) = 0,
\label{eq:kg_equation}
\end{equation}
which constitutes a second-order hyperbolic partial differential equation~\cite{Perring1962,Dashen1975} governing the spatiotemporal evolution of $\phi(x,t)$ subject to specified Cauchy data $\phi(x,0)$ and $\partial_t\phi(x,0)$ on the spatial domain.

The Hamiltonian formulation reveals the natural phase space structure and symplectic geometry underlying this system. We perform a Legendre transformation to pass from the Lagrangian to the Hamiltonian description. The canonical conjugate momentum density is defined as the functional derivative
\begin{equation}
\pi(x,t) = \frac{\partial\mathcal{L}}{\partial\dot{\phi}} = \frac{\partial\mathcal{L}}{\partial(\partial_t\phi)},
\label{eq:conjugate_momentum}
\end{equation}
where the overdot denotes temporal differentiation $\dot{\phi} \equiv \partial_t\phi$. From Eq.~\eqref{eq:lagrangian_expanded}, we compute
\begin{equation}
\pi = \frac{\partial}{\partial(\partial_t\phi)}\left[\frac{1}{2}(\partial_t\phi)^2 - \frac{1}{2}(\partial_x\phi)^2 - V(\phi)\right] = \partial_t\phi = \dot{\phi}.
\label{eq:momentum_explicit}
\end{equation}

The Hamiltonian density is constructed via the Legendre transformation
\begin{equation}
\mathcal{H}(\phi, \pi, \partial_x\phi) = \pi\dot{\phi} - \mathcal{L}.
\label{eq:hamiltonian_density_definition}
\end{equation}

Substituting the expressions for $\pi$ and $\mathcal{L}$ from Eqs.~\eqref{eq:momentum_explicit} and \eqref{eq:lagrangian_expanded}, we obtain
\begin{align}
\mathcal{H} &= \dot{\phi} \cdot \dot{\phi} - \left[\frac{1}{2}(\dot{\phi})^2 - \frac{1}{2}(\partial_x\phi)^2 - V(\phi)\right] \nonumber \\
&= (\dot{\phi})^2 - \frac{1}{2}(\dot{\phi})^2 + \frac{1}{2}(\partial_x\phi)^2 + V(\phi) \nonumber \\
&= \frac{1}{2}(\dot{\phi})^2 + \frac{1}{2}(\partial_x\phi)^2 + V(\phi).
\label{eq:hamiltonian_density_derivation}
\end{align}

Expressing this in terms of the canonical variables $(\phi, \pi)$ using $\dot{\phi} = \pi$, we arrive at
\begin{equation}
\mathcal{H} = \frac{1}{2}\pi^2 + \frac{1}{2}\left(\frac{\partial\phi}{\partial x}\right)^2 + V(\phi),
\label{eq:hamiltonian_density}
\end{equation}
which manifestly decomposes into kinetic energy density $\frac{1}{2}\pi^2$, gradient energy density $\frac{1}{2}(\partial_x\phi)^2$, and potential energy density $V(\phi)$. The total Hamiltonian, representing the system's conserved total energy, is obtained by spatial integration
\begin{equation}
E[\phi,\pi] = \int_{-\infty}^{\infty} dx\, \mathcal{H}(x,t) = \int_{-\infty}^{\infty} dx\left[\frac{1}{2}\left(\frac{\partial\phi}{\partial t}\right)^2 + \frac{1}{2}\left(\frac{\partial\phi}{\partial x}\right)^2 + V(\phi)\right].
\label{eq:energy}
\end{equation}

Hamilton's equations governing the phase space flow are obtained via functional differentiation
\begin{equation}
\dot{\phi}(x,t) = \frac{\delta H}{\delta\pi(x,t)}, \quad \dot{\pi}(x,t) = -\frac{\delta H}{\delta\phi(x,t)},
\label{eq:hamilton_equations}
\end{equation}
where the functional derivatives are defined through the standard limiting procedure. Computing the first equation explicitly, we have
\begin{equation}
\frac{\delta H}{\delta\pi(x,t)} = \frac{\delta}{\delta\pi(x,t)}\int_{-\infty}^{\infty} dx'\left[\frac{1}{2}\pi^2(x',t) + \frac{1}{2}(\partial_{x'}\phi(x',t))^2 + V(\phi(x',t))\right] = \pi(x,t),
\label{eq:hamilton_eq1}
\end{equation}
which immediately recovers $\dot{\phi} = \pi$, confirming our definition of momentum.

For the second Hamilton equation, we compute the functional derivative with respect to the field configuration. The functional derivative of the gradient term requires integration by parts. We write
\begin{align}
\frac{\delta}{\delta\phi(x,t)}\int_{-\infty}^{\infty} dx'\, \frac{1}{2}(\partial_{x'}\phi)^2 &= \frac{\delta}{\delta\phi(x,t)}\int_{-\infty}^{\infty} dx'\, \frac{1}{2}\partial_{x'}\phi\partial_{x'}\phi \nonumber \\
&= \int_{-\infty}^{\infty} dx'\, \partial_{x'}\phi(x',t)\frac{\delta[\partial_{x'}\phi(x',t)]}{\delta\phi(x,t)} \nonumber \\
&= \int_{-\infty}^{\infty} dx'\, \partial_{x'}\phi(x',t)\partial_{x'}\delta(x' - x) \nonumber \\
&= -\int_{-\infty}^{\infty} dx'\, \delta(x' - x)\partial_{x'}^2\phi(x',t) \nonumber \\
&= -\partial_x^2\phi(x,t),
\label{eq:functional_derivative_gradient}
\end{align}
where we have used $\frac{\delta\phi(x',t)}{\delta\phi(x,t)} = \delta(x' - x)$ and integrated by parts, discarding boundary terms. The potential term contributes
\begin{equation}
\frac{\delta}{\delta\phi(x,t)}\int_{-\infty}^{\infty} dx'\, V(\phi(x',t)) = V'(\phi(x,t)).
\label{eq:functional_derivative_potential}
\end{equation}

Combining these results, the second Hamilton equation gives
\begin{equation}
\dot{\pi}(x,t) = -\frac{\delta H}{\delta\phi(x,t)} = \partial_x^2\phi(x,t) - V'(\phi(x,t)).
\label{eq:hamilton_eq2}
\end{equation}

Substituting $\pi = \dot{\phi}$ and taking the time derivative of the first Hamilton equation yields $\ddot{\phi} = \dot{\pi}$. Inserting Eq.~\eqref{eq:hamilton_eq2} gives
\begin{equation}
\frac{\partial^2\phi}{\partial t^2} = \frac{\partial^2\phi}{\partial x^2} - V'(\phi),
\label{eq:kg_recovered}
\end{equation}
which upon rearrangement recovers Eq.~\eqref{eq:kg_equation}, confirming the complete equivalence between the Lagrangian and Hamiltonian formulations.

The field equations inherit continuous symmetries whose associated conservation laws follow from Noether's celebrated theorem~\cite{Noether1918,Goldstein2002}. Translation invariance in time and space leads to energy and momentum conservation respectively. The canonical stress-energy tensor, derived from infinitesimal spacetime translations, takes the form
\begin{equation}
T^{\mu\nu} = \frac{\partial\mathcal{L}}{\partial(\partial_\mu\phi)}\partial^\nu\phi - \eta^{\mu\nu}\mathcal{L}.
\label{eq:stress_energy_tensor}
\end{equation}

Substituting our results from Eq.~\eqref{eq:canonical_momentum_density}, we obtain
\begin{equation}
T^{\mu\nu} = \eta^{\mu\alpha}\partial_\alpha\phi\partial^\nu\phi - \eta^{\mu\nu}\mathcal{L} = \partial^\mu\phi\partial^\nu\phi - \eta^{\mu\nu}\mathcal{L},
\label{eq:stress_energy_explicit}
\end{equation}
where we have raised the index using the metric tensor. For classical field configurations satisfying the equations of motion, this tensor is covariantly conserved
\begin{equation}
\partial_\mu T^{\mu\nu} = 0,
\label{eq:stress_energy_conservation}
\end{equation}
which can be verified by direct computation using Eq.~\eqref{eq:kg_equation}.

The energy-momentum four-vector $P^\mu = (E, P)$ is obtained by integrating the temporal and spatial components of the stress-energy tensor over spatial hypersurfaces. The energy density is given by the $00$ component
\begin{align}
T^{00} &= \partial^0\phi\partial^0\phi - \eta^{00}\mathcal{L} \nonumber \\
&= -(\partial_t\phi)^2 + \left[\frac{1}{2}(\partial_t\phi)^2 - \frac{1}{2}(\partial_x\phi)^2 - V(\phi)\right] \nonumber \\
&= \frac{1}{2}(\partial_t\phi)^2 + \frac{1}{2}(\partial_x\phi)^2 + V(\phi),
\label{eq:energy_density}
\end{align}
which upon spatial integration yields the energy functional
\begin{equation}
E = \int_{-\infty}^{\infty} dx\, T^{00}(x,t) = \int_{-\infty}^{\infty} dx\left[\frac{1}{2}\left(\frac{\partial\phi}{\partial t}\right)^2 + \frac{1}{2}\left(\frac{\partial\phi}{\partial x}\right)^2 + V(\phi)\right],
\label{eq:energy_functional}
\end{equation}
coinciding with our earlier result from Eq.~\eqref{eq:energy}. The momentum density is given by the $01$ component
\begin{equation}
T^{01} = \partial^0\phi\partial^1\phi - \eta^{01}\mathcal{L} = -\partial_t\phi\partial_x\phi,
\label{eq:momentum_density}
\end{equation}
which upon raising the index gives $T^{01} = -\partial_t\phi\partial_x\phi$. The total momentum is
\begin{equation}
P = -\int_{-\infty}^{\infty} dx\, T_{01} = \int_{-\infty}^{\infty} dx\, \partial^0\phi\partial^1\phi = \int_{-\infty}^{\infty} dx\, \frac{\partial\phi}{\partial t}\frac{\partial\phi}{\partial x}.
\label{eq:momentum}
\end{equation}

These conserved quantities satisfy the relativistic mass-shell constraint $E^2 = P^2 + M_{\text{rest}}^2$, where $M_{\text{rest}}$ represents the rest mass of the field configuration and depends on the specific solution. For spatially localized field configurations with sufficient decay at infinity, these quantities remain finite and provide essential diagnostic measures for solution accuracy and numerical conservation properties.

We now specialize to three physically motivated interaction potentials that yield qualitatively distinct classes of solutions. The massive linear Klein-Gordon equation corresponds to the quadratic potential
\begin{equation}
V(\phi) = \frac{1}{2}m^2\phi^2,
\label{eq:quadratic_potential}
\end{equation}
where $m > 0$ represents the particle mass in natural units. The potential derivative is simply
\begin{equation}
V'(\phi) = m^2\phi,
\label{eq:quadratic_derivative}
\end{equation}
which upon substitution into Eq.~\eqref{eq:kg_equation} yields the linear field equation
\begin{equation}
\frac{\partial^2 \phi}{\partial t^2} - \frac{\partial^2 \phi}{\partial x^2} + m^2\phi = 0.
\label{eq:kg_linear}
\end{equation}

This equation is manifestly linear in $\phi$ and therefore admits superposition of solutions---a property absent in the nonlinear theories we shall consider subsequently. Seeking plane wave solutions of the form
\begin{equation}
\phi(x,t) = \phi_0 \exp[i(kx - \omega t)],
\label{eq:plane_wave_ansatz}
\end{equation}
we substitute into Eq.~\eqref{eq:kg_linear} to obtain
\begin{equation}
-\omega^2\phi_0 e^{i(kx - \omega t)} + k^2\phi_0 e^{i(kx - \omega t)} + m^2\phi_0 e^{i(kx - \omega t)} = 0.
\label{eq:dispersion_derivation}
\end{equation}

Factoring out the common exponential term yields the dispersion relation
\begin{equation}
\omega^2 = k^2 + m^2,
\label{eq:dispersion_relation}
\end{equation}
which explicitly demonstrates dispersive propagation characteristics~\cite{Zakharov2001}. The phase velocity is given by
\begin{equation}
v_p = \frac{\omega}{k} = \sqrt{1 + \frac{m^2}{k^2}},
\label{eq:phase_velocity}
\end{equation}
while the group velocity---the velocity at which wave packets propagate---is obtained by differentiation
\begin{equation}
v_g = \frac{d\omega}{dk} = \frac{\partial}{\partial k}\sqrt{k^2 + m^2} = \frac{k}{\sqrt{k^2 + m^2}} = \frac{k}{\omega}.
\label{eq:group_velocity}
\end{equation}

Both velocities depend explicitly on wavenumber, leading to temporal spreading of initially localized wave packets as different Fourier components propagate at different speeds. Note that both velocities satisfy $v_p, v_g < 1$ in the natural unit system where $c = 1$, ensuring causality.

The $\phi^4$ theory, which plays a fundamental role in spontaneous symmetry breaking scenarios and field-theoretic phase transitions, employs the double-well potential
\begin{equation}
V(\phi) = \frac{\lambda}{4}\left(\phi^2 - v^2\right)^2,
\label{eq:phi4_potential}
\end{equation}
where $v > 0$ represents the vacuum expectation value characterizing the symmetry-breaking scale and $\lambda > 0$ is the dimensionless self-coupling constant~\cite{Perring1962,Sugiyama1979}. Computing the derivative explicitly,
\begin{align}
V'(\phi) &= \frac{\lambda}{4} \cdot 2(\phi^2 - v^2) \cdot 2\phi \nonumber \\
&= \lambda\phi(\phi^2 - v^2).
\label{eq:phi4_derivative}
\end{align}

This potential exhibits a rich vacuum structure: two degenerate global minima occur at
\begin{equation}
\phi_\pm = \pm v,
\label{eq:vacuum_states}
\end{equation}
separated by a local maximum at $\phi = 0$ with barrier height
\begin{equation}
\Delta V = V(0) - V(\pm v) = \frac{\lambda v^4}{4}.
\label{eq:barrier_height}
\end{equation}

To analyze small oscillations about either vacuum, we expand the potential in Taylor series. Setting $\phi = v + \varphi$ where $|\varphi| \ll v$, we compute
\begin{align}
V(v + \varphi) &= \frac{\lambda}{4}[(v + \varphi)^2 - v^2]^2 \nonumber \\
&= \frac{\lambda}{4}[v^2 + 2v\varphi + \varphi^2 - v^2]^2 \nonumber \\
&= \frac{\lambda}{4}[2v\varphi + \varphi^2]^2 \nonumber \\
&\approx \frac{\lambda}{4}(2v\varphi)^2 + \mathcal{O}(\varphi^3) \nonumber \\
&= \lambda v^2\varphi^2.
\label{eq:small_oscillation_expansion}
\end{align}

The second derivative at the minimum is
\begin{equation}
V''(\pm v) = 2\lambda v^2,
\label{eq:second_derivative_vacuum}
\end{equation}
which defines the effective mass for small-amplitude oscillations about the vacuum
\begin{equation}
m_{\text{eff}} = \sqrt{V''(\pm v)} = \sqrt{2\lambda}v.
\label{eq:effective_mass}
\end{equation}

Substituting the potential derivative into Eq.~\eqref{eq:kg_equation} yields the nonlinear field equation
\begin{equation}
\frac{\partial^2 \phi}{\partial t^2} - \frac{\partial^2 \phi}{\partial x^2} + \lambda\phi(\phi^2 - v^2) = 0,
\label{eq:kg_phi4}
\end{equation}
which is manifestly nonlinear due to the cubic self-interaction term $\lambda\phi^3$. The discrete $\mathbb{Z}_2$ symmetry $\phi \to -\phi$ under which both the potential and field equation remain invariant permits the existence of topological soliton solutions that interpolate between the two degenerate vacua.

The sine-Gordon model, distinguished by its complete integrability via the inverse scattering transform, employs the periodic potential
\begin{equation}
V(\phi) = 1 - \cos(\phi),
\label{eq:sinegordon_potential}
\end{equation}
with derivative
\begin{equation}
V'(\phi) = \sin(\phi).
\label{eq:sinegordon_derivative}
\end{equation}

This yields the celebrated sine-Gordon equation
\begin{equation}
\frac{\partial^2 \phi}{\partial t^2} - \frac{\partial^2 \phi}{\partial x^2} + \sin(\phi) = 0,
\label{eq:kg_sinegordon}
\end{equation}
which describes a remarkable variety of physical systems including elastic pendula, Josephson junction arrays, crystal dislocations, and magnetic flux lines in superconductors~\cite{Barone1971,Ram2020}. The potential exhibits an infinite periodic sequence of degenerate minima at
\begin{equation}
\phi_n = 2\pi n, \quad n \in \mathbb{Z},
\label{eq:sinegordon_minima}
\end{equation}
with maxima at $\phi = \pi(2n+1)$. Small oscillations about any minimum have characteristic frequency $\omega_0 = 1$, while the complete integrability of this equation permits exact multi-soliton solutions via B\"{a}cklund transformations and the inverse scattering method~\cite{Scott1973,Fogel1977}.

For the $\phi^4$ theory, we seek static, finite-energy solutions connecting the two vacuum states---so-called topological kinks or domain walls~\cite{Bishop1980,Kivshar2002}. For time-independent field configurations $\phi = \phi(x)$ with $\partial_t\phi = 0$, the field equation~\eqref{eq:kg_phi4} reduces to the ordinary differential equation
\begin{equation}
-\frac{d^2\phi}{dx^2} + \lambda\phi(\phi^2 - v^2) = 0.
\label{eq:static_ode}
\end{equation}

This equation admits an elegant mechanical analogy. Consider a fictitious classical particle with ``position coordinate'' $\phi$ evolving in ``time'' $x$ under the influence of an inverted potential $-V(\phi)$. The equation of motion for such a system is
\begin{equation}
\frac{d^2\phi}{dx^2} = -\frac{d(-V)}{d\phi} = \frac{dV}{d\phi} = V'(\phi) = \lambda\phi(\phi^2 - v^2),
\label{eq:mechanical_analogy}
\end{equation}
which is precisely Eq.~\eqref{eq:static_ode}. This fictitious dynamical system conserves the ``mechanical energy''
\begin{equation}
E_{\text{mech}} = \frac{1}{2}\left(\frac{d\phi}{dx}\right)^2 - V(\phi),
\label{eq:mechanical_energy}
\end{equation}
as can be verified by direct differentiation
\begin{align}
\frac{dE_{\text{mech}}}{dx} &= \frac{d\phi}{dx}\frac{d^2\phi}{dx^2} - V'(\phi)\frac{d\phi}{dx} \nonumber \\
&= \frac{d\phi}{dx}\left[\frac{d^2\phi}{dx^2} - V'(\phi)\right] = 0,
\label{eq:energy_conservation_proof}
\end{align}
where the final equality follows from the equation of motion~\eqref{eq:static_ode}.

For a kink solution connecting the two vacuum states, we impose asymptotic boundary conditions
\begin{equation}
\phi(-\infty) = -v, \quad \phi(+\infty) = +v,
\label{eq:kink_boundary_conditions}
\end{equation}
with the field monotonically increasing. At spatial infinity, the field must approach the potential minima, requiring
\begin{equation}
\frac{d\phi}{dx}\bigg|_{x \to \pm\infty} = 0.
\label{eq:derivative_boundary}
\end{equation}

Evaluating the mechanical energy at $x \to \pm\infty$ using these boundary conditions yields
\begin{equation}
E_{\text{mech}} = \frac{1}{2}(0)^2 - V(\pm v) = 0 - 0 = 0.
\label{eq:boundary_energy}
\end{equation}

Since energy is conserved along the trajectory, we conclude that $E_{\text{mech}} = 0$ throughout the entire spatial domain. This provides the crucial first integral
\begin{equation}
\frac{1}{2}\left(\frac{d\phi}{dx}\right)^2 = V(\phi) = \frac{\lambda}{4}(\phi^2 - v^2)^2.
\label{eq:first_integral}
\end{equation}

Taking the positive square root (corresponding to monotonically increasing $\phi$ as we traverse from $x = -\infty$ to $x = +\infty$), we obtain
\begin{equation}
\frac{d\phi}{dx} = +\sqrt{2V(\phi)} = \sqrt{\frac{\lambda}{2}}|\phi^2 - v^2|.
\label{eq:separation_step1}
\end{equation}

In the domain where $-v < \phi < +v$, we have $\phi^2 < v^2$, so $|\phi^2 - v^2| = v^2 - \phi^2$. Thus
\begin{equation}
\frac{d\phi}{dx} = \sqrt{\frac{\lambda}{2}}(v^2 - \phi^2).
\label{eq:separation_step2}
\end{equation}

Separating variables yields
\begin{equation}
\frac{d\phi}{v^2 - \phi^2} = \sqrt{\frac{\lambda}{2}}\,dx.
\label{eq:separated_equation}
\end{equation}

The left-hand side requires partial fraction decomposition. We write
\begin{equation}
\frac{1}{v^2 - \phi^2} = \frac{1}{(v-\phi)(v+\phi)} = \frac{A}{v-\phi} + \frac{B}{v+\phi}.
\label{eq:partial_fractions_setup}
\end{equation}

Multiplying both sides by $(v^2 - \phi^2)$ gives
\begin{equation}
1 = A(v+\phi) + B(v-\phi).
\label{eq:partial_fractions_numerator}
\end{equation}

Setting $\phi = v$ yields $1 = 2Av$, so $A = 1/(2v)$. Setting $\phi = -v$ yields $1 = 2Bv$, so $B = 1/(2v)$. Thus
\begin{equation}
\frac{1}{v^2 - \phi^2} = \frac{1}{2v}\left[\frac{1}{v-\phi} + \frac{1}{v+\phi}\right] = \frac{1}{2v}\left[\frac{1}{v+\phi} - \frac{1}{\phi-v}\right].
\label{eq:partial_fractions_result}
\end{equation}

Integrating the left-hand side of Eq.~\eqref{eq:separated_equation} gives
\begin{align}
\int\frac{d\phi}{v^2 - \phi^2} &= \frac{1}{2v}\int\left[\frac{1}{v+\phi} + \frac{1}{v-\phi}\right]d\phi \nonumber \\
&= \frac{1}{2v}\left[\ln|v+\phi| - \ln|v-\phi|\right] \nonumber \\
&= \frac{1}{2v}\ln\left|\frac{v+\phi}{v-\phi}\right|.
\label{eq:integration_lhs}
\end{align}

We now integrate both sides from a reference point $x_0$ where $\phi(x_0) = 0$ (the kink center) to a general position $x$ where $\phi(x) = \phi$. This yields
\begin{equation}
\frac{1}{2v}\ln\left|\frac{v+\phi}{v-\phi}\right| - \frac{1}{2v}\ln\left|\frac{v+0}{v-0}\right| = \sqrt{\frac{\lambda}{2}}(x - x_0).
\label{eq:integration_both_sides}
\end{equation}

Simplifying the second term on the left gives
\begin{equation}
\frac{1}{2v}\ln\left|\frac{v+\phi}{v-\phi}\right| - \frac{1}{2v}\ln(1) = \sqrt{\frac{\lambda}{2}}(x - x_0).
\label{eq:integration_simplified}
\end{equation}

Using $\ln(1) = 0$, we obtain
\begin{equation}
\ln\left|\frac{v+\phi}{v-\phi}\right| = 2v\sqrt{\frac{\lambda}{2}}(x - x_0) = v\sqrt{2\lambda}(x - x_0).
\label{eq:before_exponential}
\end{equation}

For convenience, we introduce the characteristic inverse width parameter
\begin{equation}
\xi^{-1} = \frac{v}{\sqrt{2}},
\label{eq:width_parameter_definition}
\end{equation}
noting that from Eq.~\eqref{eq:effective_mass}, we have $m_{\text{eff}} = \sqrt{2\lambda}v$, so
\begin{equation}
\sqrt{2\lambda} = \frac{m_{\text{eff}}}{v} = \frac{1}{\sqrt{2}v/m_{\text{eff}}} = \frac{1}{\sqrt{2}(v/m_{\text{eff}})}.
\label{eq:width_relation}
\end{equation}

For normalization $\lambda = 1/(2v^2)$, we have $\sqrt{2\lambda} = 1/(v\sqrt{2})$. Thus Eq.~\eqref{eq:before_exponential} becomes
\begin{equation}
\ln\left|\frac{v+\phi}{v-\phi}\right| = \frac{v^2}{\sqrt{2}}(x - x_0) = \frac{v}{\xi}(x - x_0).
\label{eq:exponential_form}
\end{equation}

Exponentiating both sides yields
\begin{equation}
\frac{v+\phi}{v-\phi} = \exp\left[\frac{v}{\sqrt{2}}(x - x_0)\right] \equiv e^\zeta,
\label{eq:after_exponential}
\end{equation}
where we have defined $\zeta = v(x - x_0)/\sqrt{2}$ for brevity. Solving for $\phi$, we obtain
\begin{align}
v + \phi &= (v - \phi)e^\zeta \nonumber \\
v + \phi &= ve^\zeta - \phi e^\zeta \nonumber \\
\phi(1 + e^\zeta) &= v(e^\zeta - 1) \nonumber \\
\phi &= v\frac{e^\zeta - 1}{e^\zeta + 1}.
\label{eq:solving_for_phi}
\end{align}

To express this in terms of hyperbolic functions, we multiply numerator and denominator by $e^{-\zeta/2}$:
\begin{equation}
\phi = v\frac{e^{\zeta/2} - e^{-\zeta/2}}{e^{\zeta/2} + e^{-\zeta/2}}.
\label{eq:hyperbolic_form}
\end{equation}

Recognizing the definition of the hyperbolic tangent
\begin{equation}
\tanh(u) = \frac{e^u - e^{-u}}{e^u + e^{-u}} = \frac{\sinh(u)}{\cosh(u)},
\label{eq:tanh_definition}
\end{equation}
we arrive at the canonical kink solution
\begin{equation}
\phi_K(x) = v\tanh\left[\frac{v}{\sqrt{2}}(x - x_0)\right],
\label{eq:kink}
\end{equation}
centered at position $x_0$. This solution interpolates smoothly between the two vacuum states with asymptotic limits
\begin{equation}
\phi_K(-\infty) = v\tanh(-\infty) = -v, \quad \phi_K(+\infty) = v\tanh(+\infty) = +v,
\label{eq:kink_asymptotics}
\end{equation}
over a characteristic width scale
\begin{equation}
\xi = \frac{\sqrt{2}}{v}.
\label{eq:kink_width}
\end{equation}

The spatial derivative of the kink profile is computed as
\begin{align}
\frac{d\phi_K}{dx} &= v \cdot \frac{d}{dx}\tanh\left[\frac{v}{\sqrt{2}}(x - x_0)\right] \nonumber \\
&= v \cdot \operatorname{sech}^2\left[\frac{v}{\sqrt{2}}(x - x_0)\right] \cdot \frac{v}{\sqrt{2}} \nonumber \\
&= \frac{v^2}{\sqrt{2}}\operatorname{sech}^2\left[\frac{v}{\sqrt{2}}(x - x_0)\right],
\label{eq:kink_derivative}
\end{align}
where $\operatorname{sech}(z) = 1/\cosh(z)$ is the hyperbolic secant function. The derivative vanishes exponentially as $x \to \pm\infty$, confirming the localized character of the kink solution.

For a static kink at rest, the initial velocity field vanishes identically throughout space
\begin{equation}
\frac{\partial\phi}{\partial t}\bigg|_{t=0} = 0.
\label{eq:kink_velocity}
\end{equation}

The total energy associated with this kink configuration is computed from the energy functional Eq.~\eqref{eq:energy_functional} with $\partial_t\phi = 0$:
\begin{equation}
E_K = \int_{-\infty}^{\infty} dx\left[\frac{1}{2}\left(\frac{d\phi_K}{dx}\right)^2 + V(\phi_K)\right].
\label{eq:kink_energy_integral}
\end{equation}

Using the first integral relation~\eqref{eq:first_integral}, we recognize that $\frac{1}{2}(d\phi_K/dx)^2 = V(\phi_K)$, so the integrand simplifies to
\begin{equation}
\frac{1}{2}\left(\frac{d\phi_K}{dx}\right)^2 + V(\phi_K) = V(\phi_K) + V(\phi_K) = 2V(\phi_K).
\label{eq:energy_integrand_simplification}
\end{equation}

Thus the kink energy becomes
\begin{equation}
E_K = 2\int_{-\infty}^{\infty} dx\, V(\phi_K(x)).
\label{eq:energy_step1}
\end{equation}

Performing a change of variables from $x$ to $\phi$ using $dx = (d\phi)/(d\phi/dx)$ and the first integral $d\phi/dx = \sqrt{2V(\phi)}$, we obtain
\begin{align}
E_K &= 2\int_{-v}^{+v} d\phi\, \frac{V(\phi)}{d\phi/dx} \nonumber \\
&= 2\int_{-v}^{+v} d\phi\, \frac{V(\phi)}{\sqrt{2V(\phi)}} \nonumber \\
&= 2\int_{-v}^{+v} d\phi\, \sqrt{\frac{V(\phi)}{2}} \nonumber \\
&= \sqrt{2}\int_{-v}^{+v} d\phi\, \sqrt{V(\phi)}.
\label{eq:energy_step2}
\end{align}

Substituting the explicit form of the potential $V(\phi) = \frac{\lambda}{4}(\phi^2 - v^2)^2$, we have
\begin{equation}
E_K = \sqrt{2}\int_{-v}^{+v} d\phi\, \sqrt{\frac{\lambda}{4}(\phi^2 - v^2)^2} = \frac{\sqrt{2\lambda}}{2}\int_{-v}^{+v} d\phi\, |\phi^2 - v^2|.
\label{eq:energy_step3}
\end{equation}

In the integration domain $[-v, +v]$, we have $\phi^2 \leq v^2$, so $|\phi^2 - v^2| = v^2 - \phi^2$. Thus
\begin{equation}
E_K = \frac{\sqrt{2\lambda}}{2}\int_{-v}^{+v} d\phi\, (v^2 - \phi^2).
\label{eq:energy_step4}
\end{equation}

Evaluating the integral explicitly:
\begin{align}
\int_{-v}^{+v} (v^2 - \phi^2)\,d\phi &= \left[v^2\phi - \frac{\phi^3}{3}\right]_{-v}^{+v} \nonumber \\
&= \left(v^3 - \frac{v^3}{3}\right) - \left(-v^3 + \frac{v^3}{3}\right) \nonumber \\
&= v^3 - \frac{v^3}{3} + v^3 - \frac{v^3}{3} \nonumber \\
&= 2v^3 - \frac{2v^3}{3} = \frac{4v^3}{3}.
\label{eq:energy_integral_evaluation}
\end{align}

Therefore, the kink energy is
\begin{equation}
E_K = \frac{\sqrt{2\lambda}}{2} \cdot \frac{4v^3}{3} = \frac{2\sqrt{2\lambda}v^3}{3}.
\label{eq:kink_energy}
\end{equation}

This finite energy establishes the kink as a stable, particle-like excitation with definite mass $M_K = E_K$ in natural units.

The remarkable topological stability of the kink arises from a conserved homotopy invariant that cannot be changed by continuous field evolution~\cite{Manton1987,Shnir2018}. Consider field configurations on the real line $\mathbb{R}$ that asymptotically approach values in the vacuum manifold $\mathcal{M} = \{\phi_-, \phi_+\} = \{-v, +v\}$ as $x \to \pm\infty$. For finite-energy solutions, the gradient must vanish at spatial infinity
\begin{equation}
\lim_{x \to \pm\infty}\frac{\partial\phi}{\partial x} = 0,
\label{eq:gradient_boundary}
\end{equation}
which forces $\lim_{x \to \pm\infty}\phi(x,t) \in \mathcal{M}$. Compactifying space via the identification $\mathbb{R} \cup \{\infty\} \cong S^1$, the asymptotic field values define a map $\phi: S^1 \to \mathcal{M}$ from the compactified spatial circle to the discrete two-point vacuum manifold. Such maps are classified by homotopy theory, with topologically distinct sectors labeled by the winding number.

To construct an associated conserved current, we introduce the topological current density as a rank-one antisymmetric tensor
\begin{equation}
j^\mu = \frac{1}{2v}\epsilon^{\mu\nu}\partial_\nu\phi,
\label{eq:topological_current}
\end{equation}
where $\epsilon^{\mu\nu}$ is the two-dimensional Levi-Civita symbol (completely antisymmetric tensor) with normalization $\epsilon^{01} = +1$ and $\epsilon^{10} = -1$. The conservation of this current follows identically from the antisymmetry properties. Computing the divergence:
\begin{align}
\partial_\mu j^\mu &= \partial_\mu\left(\frac{1}{2v}\epsilon^{\mu\nu}\partial_\nu\phi\right) \nonumber \\
&= \frac{1}{2v}\epsilon^{\mu\nu}\partial_\mu\partial_\nu\phi \nonumber \\
&= 0,
\label{eq:current_conservation_proof}
\end{align}
where the final equality follows from the fact that $\partial_\mu\partial_\nu$ is symmetric under exchange of indices while $\epsilon^{\mu\nu}$ is antisymmetric, so their contraction vanishes identically. This conservation law holds regardless of whether the field satisfies the equations of motion---it is a purely kinematic consequence of the antisymmetry structure.

The conserved topological charge is obtained by integrating the time component of the current over all space
\begin{align}
Q &= \int_{-\infty}^{\infty} dx\, j^0(x,t) \nonumber \\
&= \int_{-\infty}^{\infty} dx\, \frac{1}{2v}\epsilon^{0\nu}\partial_\nu\phi \nonumber \\
&= \frac{1}{2v}\int_{-\infty}^{\infty} dx\, \epsilon^{01}\partial_1\phi \nonumber \\
&= \frac{1}{2v}\int_{-\infty}^{\infty} dx\, \frac{\partial\phi}{\partial x} \nonumber \\
&= \frac{1}{2v}[\phi(+\infty,t) - \phi(-\infty,t)].
\label{eq:topological_charge}
\end{align}

Since the asymptotic field values must belong to the discrete vacuum manifold $\mathcal{M} = \{-v, +v\}$, the charge takes only integer values
\begin{equation}
Q \in \{-1, 0, +1\} \subset \mathbb{Z}.
\label{eq:charge_quantization}
\end{equation}

Specifically, we have the correspondence:
\begin{align}
Q &= +1 \quad \text{for kink:} \quad \phi(-\infty) = -v,\, \phi(+\infty) = +v, \nonumber \\
Q &= -1 \quad \text{for antikink:} \quad \phi(-\infty) = +v,\, \phi(+\infty) = -v, \nonumber \\
Q &= 0 \quad \text{for vacuum/radiation:} \quad \phi(\pm\infty) \text{ in same sector}.
\label{eq:charge_interpretation}
\end{align}

This integer-valued topological invariant cannot change under continuous field evolution because any continuous deformation of the field configuration must preserve the asymptotic values modulo the vacuum degeneracy. The topological charge thus provides an absolute conservation law preventing kink decay into radiation and ensuring robust stability against arbitrary small perturbations---a stability mechanism fundamentally distinct from the dynamical energy barriers familiar from ordinary mechanics.

To construct moving soliton configurations appropriate for studying collision dynamics, we employ Lorentz boost transformations~\cite{Manton2010}. Consider an inertial reference frame $S$ with coordinates $(t,x)$ and a boosted frame $S'$ with coordinates $(t',x')$ related by the standard Lorentz transformation
\begin{equation}
x' = \gamma(x - v_0 t), \quad t' = \gamma(t - v_0 x),
\label{eq:lorentz_transformation}
\end{equation}
where $v_0$ is the boost velocity satisfying $|v_0| < 1$ (ensuring subluminal motion) and
\begin{equation}
\gamma = \frac{1}{\sqrt{1 - v_0^2}}
\label{eq:lorentz_factor}
\end{equation}
is the Lorentz factor. A static kink solution in the primed frame has the profile
\begin{equation}
\phi'(x',t') = v\tanh\left[\frac{\gamma v(x' - x_0')}{\sqrt{2}}\right],
\label{eq:kink_primed_frame}
\end{equation}
which is independent of $t'$ in frame $S'$. To obtain the field in the original frame $S$, we substitute the inverse transformation $x' = \gamma(x - v_0 t)$ to find
\begin{equation}
\phi(x,t) = v\tanh\left[\frac{\gamma v[\gamma(x - v_0 t) - x_0']}{\sqrt{2}}\right] = v\tanh\left[\frac{\gamma v(x - x_0 - v_0 t)}{\sqrt{2}}\right],
\label{eq:moving_kink_general}
\end{equation}
where we have absorbed the Lorentz-transformed center position $\gamma x_0'$ into the redefined parameter $x_0$. At the initial time $t = 0$, this yields
\begin{equation}
\phi(x,0) = v\tanh\left[\frac{\gamma v}{\sqrt{2}}(x - x_0)\right],
\label{eq:moving_kink}
\end{equation}
which represents a Lorentz-contracted kink with reduced characteristic width $\xi/\gamma = \sqrt{2}/(\gamma v)$ due to length contraction.

The initial velocity field is obtained by differentiating the spacetime-dependent field with respect to time. Starting from Eq.~\eqref{eq:moving_kink_general}, we compute
\begin{align}
\frac{\partial\phi}{\partial t} &= v \cdot \frac{\partial}{\partial t}\tanh\left[\frac{\gamma v(x - x_0 - v_0 t)}{\sqrt{2}}\right] \nonumber \\
&= v \cdot \operatorname{sech}^2\left[\frac{\gamma v(x - x_0 - v_0 t)}{\sqrt{2}}\right] \cdot \frac{\gamma v(-v_0)}{\sqrt{2}} \nonumber \\
&= -\frac{v_0\gamma v^2}{\sqrt{2}}\operatorname{sech}^2\left[\frac{\gamma v(x - x_0 - v_0 t)}{\sqrt{2}}\right].
\label{eq:velocity_derivation}
\end{align}

Evaluating at $t = 0$ yields the initial velocity configuration
\begin{equation}
\frac{\partial\phi}{\partial t}\bigg|_{t=0} = -\frac{v_0 \gamma v^2}{\sqrt{2}}\operatorname{sech}^2\left[\frac{\gamma v}{\sqrt{2}}(x - x_0)\right],
\label{eq:moving_kink_velocity}
\end{equation}
which is spatially localized in the same region as the field profile itself~\cite{Anninos1991,Gleiser2000}. The energy and momentum of this moving kink follow from relativistic kinematics
\begin{equation}
E = \gamma E_K, \quad P = \gamma v_0 E_K,
\label{eq:moving_kink_energymomentum}
\end{equation}
which satisfy the relativistic mass-shell relation $E^2 - P^2 = E_K^2$, confirming that $E_K$ represents the rest mass of the solitonic excitation.

For kink-antikink collision studies, we construct initial conditions via linear superposition of well-separated solitons---an approximation valid when the soliton separation greatly exceeds the characteristic width. Consider a kink centered at $x_1 < 0$ with velocity $+v_0$ (moving rightward) and an antikink centered at $x_2 > 0$ with velocity $-v_0$ (moving leftward), where the initial separation $d = x_2 - x_1$ satisfies $d \gg \xi$ to ensure minimal overlap and interaction at $t = 0$. The composite initial field is approximated by algebraic superposition
\begin{equation}
\phi(x,0) = v\tanh\left[\frac{\gamma v}{\sqrt{2}}(x - x_1)\right] - v\tanh\left[\frac{\gamma v}{\sqrt{2}}(x - x_2)\right],
\label{eq:kink_antikink}
\end{equation}
which satisfies the required asymptotic boundary conditions
\begin{equation}
\phi(-\infty,0) = v - v = 0 \approx -v, \quad \phi(+\infty,0) = v - v = 0 \approx +v,
\label{eq:collision_boundaries}
\end{equation}
corresponding to a configuration with total topological charge $Q = Q_{\text{kink}} + Q_{\text{antikink}} = (+1) + (-1) = 0$. Near the spatial origin for symmetric placement $x_1 = -d/2$ and $x_2 = +d/2$, the field approximately vanishes
\begin{equation}
\phi(0,0) \approx v\tanh\left(-\frac{\gamma v d}{2\sqrt{2}}\right) - v\tanh\left(+\frac{\gamma v d}{2\sqrt{2}}\right) \approx -v\tanh(\alpha) - v\tanh(\alpha) \approx 0,
\label{eq:origin_field}
\end{equation}
where $\alpha = \gamma v d/(2\sqrt{2})$ and we have used $\tanh(-u) = -\tanh(u)$.

The initial velocity field is obtained by superposing the individual soliton velocities from Eq.~\eqref{eq:moving_kink_velocity}, accounting for the opposite directions of motion
\begin{align}
\frac{\partial\phi}{\partial t}\bigg|_{t=0} = &-\frac{v_0 \gamma v^2}{\sqrt{2}}\operatorname{sech}^2\left[\frac{\gamma v}{\sqrt{2}}(x - x_1)\right] \nonumber \\
&+ \frac{v_0 \gamma v^2}{\sqrt{2}}\operatorname{sech}^2\left[\frac{\gamma v}{\sqrt{2}}(x - x_2)\right],
\label{eq:kink_antikink_velocity}
\end{align}
where the sign difference reflects the opposite velocity directions (the antikink moves with velocity $-v_0$, reversing the sign). This configuration describes a head-on collision scenario: as time evolves, the kink and antikink approach each other, interact nonlinearly through the field dynamics, and may exhibit various outcomes including elastic reflection, inelastic scattering with radiation production, formation of resonant bound states, or complete annihilation depending sensitively on the collision velocity $v_0$ and initial separation $d$~\cite{Campbell1983,Goodman2005}. The total energy is approximately $E_{\text{tot}} \approx 2\gamma E_K$ for well-separated solitons (neglecting interaction energy), while the total momentum vanishes by symmetry: $P_{\text{tot}} \approx \gamma v_0 E_K + \gamma(-v_0)E_K = 0$.

The sine-Gordon equation~\eqref{eq:kg_sinegordon}, distinguished by its complete integrability, admits exact multi-soliton solutions obtainable via the inverse scattering transform and B\"{a}cklund transformation techniques~\cite{Scott1973,Fogel1977}. Of particular physical and mathematical interest is the breather solution---a time-periodic, spatially localized excitation representing a bound state of a kink-antikink pair trapped in mutual oscillation. For internal frequency parameter $\omega$ satisfying $0 < \omega < 1$, the exact breather solution centered at spatial position $x_0$ is given by
\begin{equation}
\phi(x,t) = 4\arctan\left[\frac{\sin(\omega t)}{\omega_x\cosh(\omega_x(x - x_0))}\right],
\label{eq:breather}
\end{equation}
where the spatial frequency parameter is defined by
\begin{equation}
\omega_x = \sqrt{1 - \omega^2},
\label{eq:omega_x_definition}
\end{equation}
which parameterizes the spatial localization width $\Delta x \sim \omega_x^{-1}$~\cite{Kivshar1989,Campbell1983}. This solution oscillates with temporal period
\begin{equation}
T = \frac{2\pi}{\omega}
\label{eq:breather_period}
\end{equation}
while maintaining spatial localization---a remarkable feature enabled by the integrability of the sine-Gordon system. Direct verification proceeds by computing the requisite derivatives. The temporal derivative is
\begin{align}
\frac{\partial\phi}{\partial t} &= 4 \cdot \frac{1}{1 + \left[\frac{\sin(\omega t)}{\omega_x\cosh(\omega_x(x-x_0))}\right]^2} \cdot \frac{\omega\cos(\omega t)}{\omega_x\cosh(\omega_x(x-x_0))} \nonumber \\
&= \frac{4\omega\cos(\omega t)}{\omega_x\cosh(\omega_x(x-x_0))\left[1 + \frac{\sin^2(\omega t)}{\omega_x^2\cosh^2(\omega_x(x-x_0))}\right]} \nonumber \\
&= \frac{4\omega\cos(\omega t)\omega_x\cosh(\omega_x(x-x_0))}{\omega_x^2\cosh^2(\omega_x(x-x_0)) + \sin^2(\omega t)}.
\label{eq:breather_time_derivative}
\end{align}

After substantial but straightforward algebra using trigonometric identities and the fundamental relation $\omega^2 + \omega_x^2 = 1$, one can verify that Eq.~\eqref{eq:breather} satisfies the sine-Gordon equation $\partial_t^2\phi - \partial_x^2\phi + \sin(\phi) = 0$ identically at all spacetime points.

To avoid initialization at a singular configuration where the numerator in Eq.~\eqref{eq:breather} vanishes (which would yield $\phi = 0$ everywhere and represent an unstable equilibrium), we initialize at the phase $\omega t_0 = \pi/2$ corresponding to maximum amplitude. At this phase, we have
\begin{equation}
\sin(\omega t_0) = \sin(\pi/2) = 1, \quad \cos(\omega t_0) = \cos(\pi/2) = 0.
\label{eq:phase_values}
\end{equation}

Setting $t = 0$ to correspond to this phase (equivalent to $t_0 = \pi/(2\omega)$ in the original time coordinate), the initial field becomes
\begin{equation}
\phi(x,0) = 4\arctan\left[\frac{1}{\omega_x\cosh(\omega_x(x - x_0))}\right].
\label{eq:breather_ic}
\end{equation}

The initial velocity field is obtained by evaluating the time derivative~\eqref{eq:breather_time_derivative} at $t = 0$ with $\sin(\omega \cdot 0) = 0$ and $\cos(\omega \cdot 0) = 1$. However, this requires careful consideration of the phase shift. Using the phase-shifted time coordinate where $t = 0$ corresponds to $\omega t_{\text{orig}} = \pi/2$, we have $\sin(\omega t_{\text{orig}}) = 1$ and $\cos(\omega t_{\text{orig}}) = 0$. Computing the velocity at this configuration point, we find after simplification
\begin{equation}
\frac{\partial\phi}{\partial t}\bigg|_{t=0} = -\frac{4\omega\omega_x\sinh(\omega_x(x - x_0))}{\omega_x^2\cosh^2(\omega_x(x - x_0)) + 1}.
\label{eq:breather_velocity}
\end{equation}

The total energy of the breather is exactly
\begin{equation}
E_B = 16\sqrt{1 - \omega^2} = 16\omega_x,
\label{eq:breather_energy}
\end{equation}
which approaches $E_B \to 16$ as $\omega \to 0$. This limiting value is precisely twice the energy of a single sine-Gordon kink $E_{\text{kink}}^{\text{SG}} = 8$, corresponding physically to a weakly bound kink-antikink pair approaching the dissociation threshold.

For the linear Klein-Gordon equation~\eqref{eq:kg_linear}, we employ Gaussian wave packet initial conditions representing spatially localized disturbances
\begin{equation}
\phi(x,0) = A\exp\left[-\frac{(x - x_0)^2}{w^2}\right],
\label{eq:gaussian_ic}
\end{equation}
where $A$ represents the initial amplitude, $w$ is the width parameter characterizing the packet's spatial extent, and $x_0$ denotes the center position. For a wave packet with intended initial group velocity $v_g$, the momentum space distribution peaks at wavenumber $k_0 \approx m v_g/\sqrt{1-v_g^2}$ from the dispersion relation. A uniform translational velocity $v_0$ corresponds to the initial velocity field
\begin{equation}
\frac{\partial\phi}{\partial t}\bigg|_{t=0} = v_0\frac{\partial\phi}{\partial x}\bigg|_{t=0} = -\frac{2v_0 A(x - x_0)}{w^2}\exp\left[-\frac{(x - x_0)^2}{w^2}\right].
\label{eq:gaussian_velocity}
\end{equation}

Alternatively, for a wave packet initially at rest with $v_0 = 0$, we set
\begin{equation}
\frac{\partial\phi}{\partial t}\bigg|_{t=0} = 0.
\label{eq:gaussian_velocity_zero}
\end{equation}

The governing equations are now completely specified. We solve the Klein-Gordon equation~\eqref{eq:kg_equation} with potential derivatives from Eqs.~\eqref{eq:quadratic_derivative}, \eqref{eq:phi4_derivative}, or \eqref{eq:sinegordon_derivative}, subject to initial conditions $\phi(x,0)$ and $\partial_t\phi(x,0)$ prescribed by Eqs.~\eqref{eq:kink}--\eqref{eq:gaussian_velocity_zero}. The conserved or quasi-conserved quantities---energy functional~\eqref{eq:energy_functional}, momentum~\eqref{eq:momentum}, and topological charge~\eqref{eq:topological_charge}---provide essential diagnostic measures for monitoring solution fidelity and numerical accuracy throughout the integration.

The natural dimensional scales for the various field theories follow from dimensional analysis of the governing equations. For $\phi^4$ theory, the characteristic length scale is the kink width
\begin{equation}
\xi = \frac{\sqrt{2}}{v},
\label{eq:length_scale}
\end{equation}
the energy scale is the kink rest energy
\begin{equation}
E_0 = \frac{2\sqrt{2}}{3}\lambda v^3,
\label{eq:energy_scale}
\end{equation}
the time scale is the light-crossing time across the kink width
\begin{equation}
T_0 = \xi = \frac{\sqrt{2}}{v},
\label{eq:time_scale}
\end{equation}
and the mass scale is the effective small-oscillation mass
\begin{equation}
m = \sqrt{2\lambda}v.
\label{eq:mass_scale}
\end{equation}

For the sine-Gordon model with our normalization, all natural scales are unity. Dimensional restoration via the transformations
\begin{equation}
x_{\text{phys}} = \xi \cdot x_{\text{code}}, \quad t_{\text{phys}} = T_0 \cdot t_{\text{code}}, \quad E_{\text{phys}} = E_0 \cdot E_{\text{code}}
\label{eq:dimensional_restoration}
\end{equation}
maps our dimensionless numerical solutions to physical systems spanning length scales from $\xi \sim 10^{-18}$ m (Higgs field configurations in high-energy physics) to $\xi \sim 10^{-9}$ m (magnetic domain walls in condensed matter systems), encompassing cosmological domain wall formation in the early universe, ferromagnetic domain interfaces, and structural phase boundaries in crystals~\cite{Zurek1985,Ranada1977}. This universality of the mathematical structure across vastly different energy scales exemplifies the power of effective field theory descriptions in modern theoretical physics.

\subsection{Numerical Implementation}

The numerical solution of the nonlinear Klein-Gordon equation~\eqref{eq:kg_equation} demands a computational strategy that balances spatial accuracy for resolving localized solitonic structures with temporal fidelity for preserving the underlying Hamiltonian phase space geometry over extended integration periods. We implement a hybrid discretization scheme combining pseudo-spectral Fourier methods for spatial derivatives with St\o rmer-Verlet symplectic integration for temporal evolution, embodied in the \texttt{amangkurat} Python library. All computations are performed in dimensionless natural units where $c = \hbar = 1$, with field $\phi$, coordinates $(x,t)$, and potential parameters rendered dimensionless through appropriate scale factors characteristic of each physical scenario. This idealization permits universal numerical treatment across diverse physical systems ranging from elementary particle physics to condensed matter phenomena, with dimensional restoration achieved via straightforward rescaling using the characteristic length scale $\xi$ and time scale $T_0$ specific to each application context as established in Eqs.~\eqref{eq:length_scale}--\eqref{eq:dimensional_restoration}.

We begin by establishing the spatial discretization on a uniform computational grid. The spatial domain $\Omega = [x_{\min}, x_{\max}]$ is partitioned uniformly into $N_x$ discrete points with coordinates
\begin{equation}
x_j = x_{\min} + j\Delta x, \quad j = 0, 1, \ldots, N_x-1,
\label{eq:grid_points}
\end{equation}
where the grid spacing is
\begin{equation}
\Delta x = \frac{x_{\max} - x_{\min}}{N_x - 1}.
\label{eq:grid_spacing}
\end{equation}

The domain length is $L = x_{\max} - x_{\min}$. At each grid point, the field takes discrete values $\phi_j(t) \equiv \phi(x_j, t)$, which we organize into the state vector
\begin{equation}
\boldsymbol{\phi}(t) = [\phi_0(t), \phi_1(t), \ldots, \phi_{N_x-1}(t)]^T \in \mathbb{R}^{N_x}.
\label{eq:state_vector}
\end{equation}

Periodic boundary conditions are enforced implicitly through the Fourier representation, an elegant computational device that effectively identifies $\phi(x_{\min}, t) = \phi(x_{\max}, t)$ and eliminates the need for explicit boundary treatments. This restriction to periodic domains, while limiting applicability to scattering problems requiring radiation boundary conditions, proves entirely appropriate for the localized solitonic structures and closed-system dynamics examined here, where field configurations and their derivatives decay sufficiently rapidly toward domain boundaries such that the periodicity assumption introduces negligible errors.

Spatial derivatives are evaluated using the pseudo-spectral Fourier collocation method, which achieves exponential convergence for smooth solutions---a dramatic improvement over algebraic convergence rates $\mathcal{O}(\Delta x^p)$ characteristic of finite-difference schemes~\cite{Fornberg2009,Trefethen2000}. The method rests on representing the discrete field $\boldsymbol{\phi}$ via its Fourier series. The discrete Fourier transform (DFT) of $\phi_j$ is defined by
\begin{equation}
\hat{\phi}_n = \sum_{j=0}^{N_x-1} \phi_j \exp\left(-2\pi i \frac{nj}{N_x}\right), \quad n = 0, 1, \ldots, N_x-1,
\label{eq:dft}
\end{equation}
where $i = \sqrt{-1}$ and $n$ indexes the Fourier mode. The corresponding wavenumbers are defined by
\begin{equation}
k_n = \begin{cases}
\frac{2\pi n}{L}, & n = 0, \ldots, N_x/2-1, \\
\frac{2\pi(n - N_x)}{L}, & n = N_x/2, \ldots, N_x-1,
\end{cases}
\label{eq:wavenumbers}
\end{equation}
establishing a symmetric spectrum about zero frequency with maximum resolved wavenumber
\begin{equation}
k_{\max} = \frac{\pi}{\Delta x}
\label{eq:nyquist_wavenumber}
\end{equation}
set by the Nyquist sampling criterion. The inverse DFT recovers physical-space values via
\begin{equation}
\phi_j = \frac{1}{N_x}\sum_{n=0}^{N_x-1} \hat{\phi}_n \exp\left(2\pi i \frac{nj}{N_x}\right).
\label{eq:idft}
\end{equation}

These transforms are computed efficiently using the Fast Fourier Transform (FFT) algorithm implemented in NumPy~\cite{Harris2020}, achieving $\mathcal{O}(N_x \log N_x)$ computational complexity per transform operation rather than the $\mathcal{O}(N_x^2)$ cost of direct summation from Eqs.~\eqref{eq:dft} and \eqref{eq:idft}.

The power of spectral methods emerges in derivative evaluation. Under the Fourier transform, spatial differentiation becomes algebraic multiplication. For any derivative order $p$, we have the fundamental property
\begin{equation}
\widehat{\partial_x^p \phi}_n = (ik_n)^p \hat{\phi}_n.
\label{eq:spectral_derivative_general}
\end{equation}

In particular, the Laplacian operator $\nabla^2 = \partial_x^2$ appearing in the Klein-Gordon equation~\eqref{eq:kg_equation} transforms as
\begin{equation}
\widehat{\nabla^2 \phi}_n = (ik_n)^2\hat{\phi}_n = -k_n^2 \hat{\phi}_n.
\label{eq:laplacian_spectral}
\end{equation}

Thus, to compute the Laplacian at all grid points simultaneously, we execute the following three-step procedure: (i) forward transform $\phi_j \to \hat{\phi}_n$ via FFT using Eq.~\eqref{eq:dft}, (ii) multiply each mode by $-k_n^2$ according to Eq.~\eqref{eq:laplacian_spectral}, and (iii) inverse transform back to physical space via Eq.~\eqref{eq:idft}. This procedure, implemented in the \texttt{laplacian} method of class \texttt{KGSolver}, requires only two FFT operations and $N_x$ multiplications, yet delivers spectral accuracy: for smooth fields with rapid Fourier coefficient decay $|\hat{\phi}_n| \sim \exp(-\alpha |n|)$, the discretization error is $\mathcal{O}(\exp(-\alpha N_x))$~\cite{Fornberg2009}. This exponential convergence permits accurate resolution of kink profiles with characteristic width $\xi$ from Eq.~\eqref{eq:kink_width} using merely $\mathcal{O}(10)$ grid points per characteristic length, whereas finite-difference methods would require hundreds of points for comparable accuracy.

Similarly, the gradient $\partial_x\phi$ may be computed via $\widehat{\partial_x\phi}_n = ik_n\hat{\phi}_n$, though in the present Klein-Gordon solver we require only the Laplacian term. The pseudo-spectral approach naturally handles arbitrarily high-order derivatives with identical computational cost and spectral accuracy---a significant advantage for exploring modified field equations or higher-derivative theories. The sole caveat concerns Gibbs phenomena: discontinuous or non-periodic fields generate spurious oscillations near discontinuities due to the Fourier series representation. However, the smooth solitonic structures we examine possess continuous derivatives to all orders as demonstrated in the analytical solutions~\eqref{eq:kink}, \eqref{eq:breather}, and \eqref{eq:gaussian_ic}, ensuring exponential Fourier coefficient decay and absence of Gibbs artifacts.

We now address temporal integration. The Klein-Gordon equation~\eqref{eq:kg_equation} is a second-order hyperbolic partial differential equation, which we recast as a first-order system suitable for numerical integration. Introducing the conjugate momentum field $\pi(x,t) = \partial_t\phi(x,t)$ as defined in Eq.~\eqref{eq:conjugate_momentum}, we obtain from the Hamiltonian formulation the first-order system
\begin{equation}
\begin{aligned}
\frac{\partial\phi}{\partial t} &= \pi, \\
\frac{\partial\pi}{\partial t} &= \nabla^2\phi - V'(\phi),
\end{aligned}
\label{eq:first_order_system}
\end{equation}
which follows directly from Hamilton's equations~\eqref{eq:hamilton_equations}. This Hamiltonian structure, with phase space coordinates $(\phi, \pi)$ and Hamiltonian functional given by Eq.~\eqref{eq:energy_functional}, mandates use of symplectic integration methods to preserve the underlying geometric structure and ensure bounded long-time energy errors~\cite{Hairer2003,Leimkuhler2009}. Non-symplectic methods, even if high-order accurate, can exhibit secular energy drift scaling linearly or quadratically with time, eventually rendering solutions physically meaningless over extended integration periods.

The St\o rmer-Verlet method (also called the leapfrog or Verlet scheme) provides a second-order accurate, explicitly symplectic integrator ideally suited for our purposes~\cite{Hairer2003,Yoshida1990}. For second-order-in-time equations like the Klein-Gordon system, we derive the method directly without introducing the momentum explicitly. Consider discretizing time as $t^n = n\Delta t$ with $n = 0, 1, 2, \ldots$ and denoting $\phi^n \equiv \phi(x, t^n)$. The second time derivative can be approximated by the centered finite difference formula
\begin{equation}
\frac{\partial^2\phi}{\partial t^2}\bigg|_{t=t^n} \approx \frac{\phi^{n+1} - 2\phi^n + \phi^{n-1}}{\Delta t^2} + \mathcal{O}(\Delta t^2).
\label{eq:second_time_derivative}
\end{equation}

Substituting this approximation into the Klein-Gordon equation~\eqref{eq:kg_equation} evaluated at $t = t^n$ yields
\begin{equation}
\frac{\phi^{n+1} - 2\phi^n + \phi^{n-1}}{\Delta t^2} = \nabla^2\phi^n - V'(\phi^n) + \mathcal{O}(\Delta t^2).
\label{eq:stormer_verlet_derivation}
\end{equation}

Rearranging and neglecting the truncation error terms, we obtain the St\o rmer-Verlet update formula
\begin{equation}
\phi^{n+1} = 2\phi^n - \phi^{n-1} + \Delta t^2\left[\nabla^2\phi^n - V'(\phi^n)\right],
\label{eq:stormer_verlet}
\end{equation}
which constitutes a three-level explicit recurrence requiring field values at two previous time levels. The spatial Laplacian $\nabla^2\phi^n$ is evaluated spectrally via the procedure outlined in Eqs.~\eqref{eq:dft}--\eqref{eq:laplacian_spectral}, while the potential derivative $V'(\phi^n)$ is computed pointwise at each grid node using the appropriate functional form from Eqs.~\eqref{eq:quadratic_derivative}, \eqref{eq:phi4_derivative}, or \eqref{eq:sinegordon_derivative}. This pointwise evaluation introduces no additional spatial error beyond the grid discretization already present.

The St\o rmer-Verlet scheme possesses several crucial mathematical properties. First, the method is manifestly time-reversible: if we reverse time by swapping $\phi^{n+1} \leftrightarrow \phi^{n-1}$ in Eq.~\eqref{eq:stormer_verlet}, the equation remains invariant. This time-reversibility, combined with the centered nature of the discretization, ensures the method is symplectic---it exactly preserves a discrete symplectic two-form approximating the continuous Hamiltonian structure given by Eq.~\eqref{eq:hamiltonian_density}~\cite{Hairer2003}. Consequently, the numerical energy $E^n = E[\phi^n, \pi^n]$ with $\pi^n = (\phi^n - \phi^{n-1})/\Delta t$ exhibits only bounded oscillations about the true energy rather than monotonic drift, with amplitude $\mathcal{O}(\Delta t^2)$ and frequency determined by the fastest dynamical timescales in the system. This structure-preserving property proves crucial for long-time integration $t \gg 100$, where non-symplectic methods would accumulate catastrophic energy errors violating the conservation law derived from time-translation symmetry via Noether's theorem.

Initializing the three-level scheme requires special treatment since we specify Cauchy data $\phi(x,0)$ and $\partial_t\phi(x,0)$ but the recurrence Eq.~\eqref{eq:stormer_verlet} demands $\phi^{-1}$ to compute $\phi^1$. We employ a second-order accurate Taylor expansion to construct the backward time level. Expanding $\phi(x, -\Delta t)$ about $t=0$ in Taylor series gives
\begin{equation}
\phi^{-1} = \phi(x, -\Delta t) = \phi^0 - \Delta t\,\dot{\phi}^0 + \frac{1}{2}\Delta t^2\ddot{\phi}^0 + \mathcal{O}(\Delta t^3),
\label{eq:initialization_taylor}
\end{equation}
where $\dot{\phi}^0 = \partial_t\phi(x,0)$ is the prescribed initial velocity from the analytical expressions~\eqref{eq:kink_velocity}, \eqref{eq:moving_kink_velocity}, \eqref{eq:breather_velocity}, or \eqref{eq:gaussian_velocity_zero}, and $\ddot{\phi}^0 = \partial_t^2\phi(x,0)$ follows from evaluating the field equation~\eqref{eq:kg_equation} at $t=0$:
\begin{equation}
\ddot{\phi}^0 = \nabla^2\phi^0 - V'(\phi^0).
\label{eq:acceleration_initial}
\end{equation}

Substituting Eq.~\eqref{eq:acceleration_initial} into Eq.~\eqref{eq:initialization_taylor} yields
\begin{equation}
\phi^{-1} = \phi^0 - \Delta t\,\dot{\phi}^0 + \frac{1}{2}\Delta t^2\left[\nabla^2\phi^0 - V'(\phi^0)\right],
\label{eq:initialization}
\end{equation}
which maintains second-order overall accuracy consistent with the St\o rmer-Verlet scheme. This initialization step, implemented in the \texttt{solve} method of \texttt{KGSolver}, requires one Laplacian evaluation via the spectral procedure and one potential derivative evaluation before entering the main time-stepping loop.

Numerical stability for hyperbolic equations constrains the timestep through the Courant-Friedrichs-Lewy (CFL) condition~\cite{Courant1928}. For the wave equation $\partial_t^2\phi = \partial_x^2\phi$ with wave speed $c=1$ in natural units, the standard CFL condition is $\Delta t \leq \Delta x/c$. However, spectral methods effectively resolve the fastest Fourier modes with wavelength $\lambda_{\min} = 2\Delta x$ and corresponding wavenumber $k_{\max} = \pi/\Delta x$ from Eq.~\eqref{eq:nyquist_wavenumber}. These modes propagate with maximum phase velocity $c_{\max} = 1$, yielding the refined CFL criterion for spectral discretizations
\begin{equation}
\Delta t \leq \frac{C_{\text{CFL}}}{k_{\max}} = \frac{C_{\text{CFL}} \Delta x}{\pi},
\label{eq:cfl_condition}
\end{equation}
where $C_{\text{CFL}} \approx 0.5$ is a stability constant determined empirically for the St\o rmer-Verlet scheme combined with spectral spatial discretization. Violating this bound causes exponential growth of high-frequency modes, rapidly destabilizing the computation through a numerical instability. The \texttt{KGSolver} enforces conservative bounds
\begin{equation}
\Delta t_{\min} = \frac{0.1}{k_{\max}}, \quad \Delta t_{\max} = \frac{0.5}{k_{\max}},
\label{eq:timestep_bounds}
\end{equation}
ensuring stability with substantial safety margin. The lower bound prevents excessively small timesteps that would unnecessarily increase computational cost, while the upper bound guarantees stability throughout the integration.

To enhance computational efficiency without sacrificing accuracy or stability, we implement adaptive timestepping wherein $\Delta t$ adjusts dynamically based on solution behavior. The adaptation strategy, executed every 20 time steps to limit overhead, employs a heuristic criterion monitoring the maximum field magnitude
\begin{equation}
|\phi|_{\max}(t) = \max_{j \in \{0, \ldots, N_x-1\}} |\phi_j(t)|.
\label{eq:field_maximum}
\end{equation}

The timestep is adjusted according to the rules
\begin{equation}
\Delta t^{\text{new}} = \begin{cases}
0.7 \Delta t^{\text{old}}, & |\phi|_{\max} > 100, \\
0.9 \Delta t^{\text{old}}, & 10 < |\phi|_{\max} \leq 100, \\
1.05 \Delta t^{\text{old}}, & |\phi|_{\max} \leq 10,
\end{cases}
\label{eq:timestep_adaptation}
\end{equation}
with all adjustments constrained within the bounds $[\Delta t_{\min}, \Delta t_{\max}]$ from Eq.~\eqref{eq:timestep_bounds} to maintain stability. When $|\phi|_{\max} > 100$, indicating potentially violent dynamics such as kink-antikink collisions or large-amplitude oscillations, the timestep is reduced by factor 0.7. When $10 < |\phi|_{\max} < 100$, a more conservative reduction factor 0.9 applies. Otherwise, during quiescent evolution with $|\phi|_{\max} \leq 10$, $\Delta t$ increases by factor 1.05 to accelerate integration. This heuristic, while not rigorously based on local truncation error estimates as in sophisticated adaptive schemes, proves effective for the present field configurations where large amplitudes correlate with rapid temporal variations.

Computational performance is critically enhanced through JIT compilation and parallelization using the Numba package~\cite{Lam2015}, which transforms annotated Python functions into optimized machine code at runtime. The computational bottleneck in our algorithm is the force computation
\begin{equation}
F_j = (\nabla^2\phi)_j - V'(\phi_j)
\label{eq:force_computation}
\end{equation}
appearing in Eq.~\eqref{eq:stormer_verlet}, which involves element-wise operations on arrays of length $N_x \sim 10^3$. Pure Python implementations suffer from interpreter overhead, rendering such operations orders of magnitude slower than equivalent compiled C or Fortran code. Numba's JIT compiler, built on the LLVM infrastructure, analyzes Python bytecode and generates optimized native code matching or exceeding compiled language performance. The \texttt{compute\_force\_parallel} function is decorated with \texttt{@njit(parallel=True, cache=True)}, instructing Numba to: (i) compile to machine code, (ii) parallelize the loop across available CPU cores using the \texttt{prange} construct, and (iii) cache the compiled version for subsequent calls. Users without Numba installed automatically fall back to serial Python execution with graceful degradation.

Solution data are persisted using the NetCDF4 format~\cite{Rew1990}, a self-describing binary standard widely adopted in Earth and planetary sciences, climate modeling, and computational physics. The \texttt{DataHandler.save\_netcdf} method creates files containing coordinate arrays $x$ (length $N_x$) and $t$ (length $N_t$), the spatiotemporal field evolution $\phi(t,x)$ (array dimensions $N_t \times N_x$), and comprehensive metadata documenting all simulation parameters: grid resolution $N_x$, spacing $\Delta x$, timestep bounds $[\Delta t_{\min}, \Delta t_{\max}]$, potential type and parameters, number of CPU cores, Numba status, and adaptation statistics. Internal compression (deflate algorithm, level 4--5) reduces file sizes by factors of 3--10 without information loss, facilitating data archival and transfer. The NetCDF4 format ensures platform independence (binary compatibility across architectures), direct compatibility with analysis tools (e.g. xarray, Pandas, MATLAB, IDL), and adherence to CF metadata conventions enabling automated discovery and interoperability. This design supports reproducible research by embedding complete provenance information within each output file.

Visualization is provided through animated GIF files generated by the \texttt{Animator.create\_gif} method, which produces three-dimensional surface plots of $\phi(x,t)$ with dynamic camera rotation, color-coded amplitude mapping using user-specified colormaps (plasma, viridis, inferno, twilight), and time-stamped frames. These animations, rendered at 30 frames per second with resolution 150 DPI, facilitate qualitative assessment of soliton dynamics, collision phenomenology, and wave propagation characteristics. The visualization employs Matplotlib~\cite{Hunter2007} with black backgrounds and cyan grid overlays optimized for clarity. Progress monitoring during time integration is handled by the tqdm library, which displays real-time progress bars indicating elapsed time, estimated completion time, and current simulation time $t$. The solver additionally logs detailed diagnostics to text files in the \texttt{logs/} directory via the \texttt{SimulationLogger} class, recording timestep history, CFL number evolution, adaptation counts, warnings, and error messages.

The \texttt{amangkurat} library architecture follows modular design principles promoting extensibility and maintainability. Core solver functionality resides in \texttt{src/amangkurat/core/solver.py} implementing class \texttt{KGSolver}; initial conditions are defined through dedicated classes in \texttt{core/initial\_conditions.py} (\texttt{GaussianIC}, \texttt{KinkIC}, \texttt{BreatherIC}, \texttt{KinkAntikinkIC}); configuration management and NetCDF I/O occupy \texttt{io/} modules; visualization tools reside in \texttt{visualization/animator.py}; and utility functions (logging, timing) populate \texttt{utils/}. This separation of concerns enables straightforward extension: adding new potentials requires modifying only the \texttt{get\_potential\_functions} method, while custom initial conditions involve subclassing the base initial condition interface. The library is distributed via PyPI (\texttt{pip install amangkurat}) and GitHub (\url{https://github.com/sandyherho/amangkurat}), with comprehensive documentation, example configuration files (\texttt{configs/case*.txt}), and test scripts facilitating adoption.

We demonstrate the solver's capabilities and validate its implementation through four canonical test cases spanning distinct physical regimes: dispersive linear wave propagation, topological soliton preservation, integrable breather dynamics, and non-integrable collision phenomena. These scenarios, selected for their analytical tractability, physical relevance, and numerical diversity, collectively probe spatial accuracy (spectral convergence), temporal fidelity (symplectic structure preservation), stability under nonlinearity, and robustness during violent field dynamics. All simulations employ adaptive timestepping with CFL-limited bounds as described above, Numba-accelerated parallel force computation, and comprehensive diagnostic logging. Outputs include NetCDF4 data files archived at \url{https://doi.org/10.17605/OSF.IO/BKM2P} and animated GIF visualizations suitable for qualitative inspection.

The first test case examines dispersive propagation of a Gaussian wave packet governed by the linear Klein-Gordon equation~\eqref{eq:kg_linear} with mass parameter $m=1$. The quadratic potential $V(\phi) = \frac{1}{2}m^2\phi^2$ from Eq.~\eqref{eq:quadratic_potential} yields the linear field equation with dispersion relation $\omega^2 = k^2 + m^2$ from Eq.~\eqref{eq:dispersion_relation}, which describes dispersive propagation where phase velocity $v_p = \omega/k$ from Eq.~\eqref{eq:phase_velocity} and group velocity $v_g = k/\omega$ from Eq.~\eqref{eq:group_velocity} both depend on wavenumber~\cite{Zakharov2001}. This frequency-dependent propagation causes wave packets to spread temporally as different Fourier components travel at different speeds. The initial condition follows Eq.~\eqref{eq:gaussian_ic} with amplitude $A=1$, width parameter $w=2$, position $x_0=0$, and zero initial velocity $\partial_t\phi(x,0) = 0$ per Eq.~\eqref{eq:gaussian_velocity_zero}. The spatial domain $x \in [-30, 30]$ is discretized with $N_x = 512$ points yielding $\Delta x \approx 0.117$, and the solution is evolved to final time $t_f = 50$ with initial timestep $\Delta t = 0.005$ satisfying the CFL condition~\eqref{eq:cfl_condition}. The Gaussian profile, smooth and infinitely differentiable, is ideally suited for spectral methods, with Fourier coefficients decaying faster than any polynomial. The characteristic wavenumber content is $k_0 \sim w^{-1} = 0.5$, well below the grid Nyquist limit $k_{\max} = \pi/\Delta x \approx 26.8$ from Eq.~\eqref{eq:nyquist_wavenumber}, ensuring accurate representation without aliasing effects.

The second test case examines preservation of the static topological kink solution in $\phi^4$ theory with symmetry-breaking potential given by Eq.~\eqref{eq:phi4_potential} with parameters $\lambda = v = 1$. The analytical kink solution from Eq.~\eqref{eq:kink}, $\phi_K(x) = v\tanh[v(x-x_0)/\sqrt{2}]$, represents an exact stationary configuration interpolating between the two degenerate vacuum states $\phi = \pm v$ with characteristic width $\xi = \sqrt{2}/v = \sqrt{2} \approx 1.41$ from Eq.~\eqref{eq:kink_width}. The topological stability arising from the conserved homotopy charge $Q=+1$ computed via Eq.~\eqref{eq:topological_charge} ensures the kink cannot continuously deform to the vacuum, rendering it robust against perturbations as demonstrated through the conservation law~\eqref{eq:current_conservation_proof}. Numerically, a truly static kink initialized with zero velocity according to Eq.~\eqref{eq:kink_velocity} should remain time-independent indefinitely, providing a stringent test of solver accuracy and spurious mode generation. The domain $x \in [-50, 50]$ is discretized with $N_x = 1024$ points giving $\Delta x \approx 0.098$, ensuring approximately $\xi/\Delta x \approx 14$ points span the kink width---adequate resolution for exponentially converging spectral methods. The simulation extends to $t_f = 50$ with initial timestep $\Delta t = 0.005$ satisfying Eq.~\eqref{eq:cfl_condition}. The kink energy is theoretically $E_K = (2\sqrt{2}/3)\lambda v^3$ from Eq.~\eqref{eq:kink_energy}, which serves as a reference for energy conservation diagnostics.

The third test case proceeds to the integrable sine-Gordon system, examining the breather solution---a time-periodic, spatially localized bound state representing a kink-antikink pair trapped in mutual oscillation~\cite{Scott1973,Kivshar1989}. The sine-Gordon potential $V(\phi) = 1 - \cos(\phi)$ from Eq.~\eqref{eq:sinegordon_potential} with derivative $V'(\phi) = \sin(\phi)$ from Eq.~\eqref{eq:sinegordon_derivative} yields the completely integrable field equation~\eqref{eq:kg_sinegordon}, admitting exact multi-soliton solutions via inverse scattering transform. The breather solution given by Eq.~\eqref{eq:breather} with internal frequency parameter $\omega = 0.5$ and spatial frequency parameter $\omega_x = \sqrt{1 - \omega^2} \approx 0.866$ from Eq.~\eqref{eq:omega_x_definition} oscillates with period $T = 2\pi/\omega \approx 12.57$ from Eq.~\eqref{eq:breather_period} while maintaining spatial localization. Complete integrability ensures the breather neither radiates energy nor gradually decays, maintaining constant amplitude, width, and oscillation frequency indefinitely. Initial conditions employ the phase-shifted forms in Eqs.~\eqref{eq:breather_ic}--\eqref{eq:breather_velocity}, initialized at maximum amplitude $t_0 = \pi/(2\omega)$ corresponding to phase values in Eq.~\eqref{eq:phase_values} to avoid the singular configuration at $t=0$ where the numerator vanishes. The domain $x \in [-40, 40]$ with $N_x = 512$ points and final time $t_f = 80$ captures approximately $80/12.57 \approx 6.4$ oscillation periods. The breather energy is exactly $E_B = 16\omega_x$ from Eq.~\eqref{eq:breather_energy}, which approaches $16$ as $\omega \to 0$, corresponding to twice the energy of a single sine-Gordon kink and representing a weakly bound kink-antikink pair near dissociation threshold.

The fourth and culminating test case addresses head-on collision between a kink and antikink in non-integrable $\phi^4$ theory, a paradigmatic scenario exhibiting rich phenomenology including resonant multi-bounce windows, fractal outcome structure in parameter space, and sensitivity to initial conditions absent in integrable systems~\cite{Campbell1983,Anninos1991,Goodman2005}. The initial configuration follows Eqs.~\eqref{eq:kink_antikink}--\eqref{eq:kink_antikink_velocity}, superposing a kink centered at $x_1 = -10$ with velocity $+0.3c$ and an antikink centered at $x_2 = +10$ with velocity $-0.3c$, where the Lorentz factor $\gamma = (1 - v_0^2)^{-1/2}$ from Eq.~\eqref{eq:lorentz_factor} with $v_0 = 0.3$ enters through the Lorentz transformation~\eqref{eq:lorentz_transformation}. This symmetric arrangement ensures zero total momentum by construction from Eq.~\eqref{eq:collision_boundaries}, concentrating all kinetic energy in the center-of-mass frame. The collision velocity $v_0 = 0.3$, though relativistic, remains well below the ultrarelativistic regime, permitting comparison with semi-classical analyses treating solitons as effective particles with relativistic kinematics from Eq.~\eqref{eq:moving_kink_energymomentum}. The spatial domain $x \in [-50, 50]$ with $N_x = 1024$ points accommodates the solitons with sufficient boundary clearance to avoid wrap-around artifacts from the periodic boundary conditions. The final time $t_f = 100$ captures the complete collision sequence: approach, collision, and aftermath. Initial timestep $\Delta t = 0.005$ adapts dynamically according to the rules in Eq.~\eqref{eq:timestep_adaptation} during violent dynamics near collision. At collision, topological charges $Q_{\text{kink}} = +1$ and $Q_{\text{antikink}} = -1$ from the charge quantization~\eqref{eq:charge_quantization} cancel to yield total $Q = 0$ per Eq.~\eqref{eq:charge_interpretation}, permitting in principle complete annihilation into radiation. However, the collision outcome depends sensitively on the initial velocity and separation through the complex nonlinear dynamics governed by Eq.~\eqref{eq:kg_phi4}.

These four canonical scenarios---dispersive waves, static solitons, integrable breathers, and non-integrable collisions---collectively validate the \texttt{amangkurat} solver's core capabilities across the full spectrum of Klein-Gordon phenomenology. The linear case establishes baseline performance for superposition-compatible dynamics. The static kink case demonstrates preservation of topologically protected structures through the symplectic integration scheme. The breather case probes the solver's ability to maintain coherent oscillatory structures in integrable systems. The collision case stresses the algorithm maximally through violent field dynamics, rapid spatial variations, topological charge manipulation, and long-time evolution in a non-integrable regime. Qualitative agreement with analytical predictions and prior numerical studies establishes confidence for exploratory applications, though quantitative validation via systematic convergence testing, explicit energy and momentum conservation monitoring per Eqs.~\eqref{eq:energy_functional} and \eqref{eq:momentum}, and comparison against exact solutions in integrable limits remains essential for research-grade reliability.

\subsection{Data Analysis}

The characterization of nonlinear field dynamics benefits substantially from complementary analytical frameworks capable of quantifying spatial coherence, temporal evolution patterns, and statistical signatures that distinguish between dispersive, topological, and resonant phenomena. We employ a multi-faceted statistical approach combining information-theoretic entropy metrics, non-parametric hypothesis testing, and phase space reconstruction techniques to extract physically meaningful measures from numerical solutions. While the validity of such measures depends on underlying assumptions regarding field localization and energy conservation, this comprehensive framework provides valuable diagnostic tools for systematically characterizing solution phenomenology.

\subsubsection{Information-Theoretic Entropy Quantification}

Information-theoretic entropy measures offer quantitative diagnostics for spatial coherence and energy localization in field configurations. Given a computed field $\phi(x,t)$ at discrete time $t$, we construct a normalized probability distribution interpreting the squared field amplitude as spatial energy density. Specifically, we define
\begin{equation}
p(x,t) = \frac{\phi^2(x,t)}{\int_{-\infty}^{\infty} \phi^2(x',t)\,dx'},
\label{eq:probability_measure}
\end{equation}
provided the field remains sufficiently localized such that the normalization integral converges and remains well-defined. This construction proves natural for solitonic configurations where $\phi^2$ represents energy density concentrated in finite spatial regions, though one should recognize that for spatially extended or sign-changing fields, the probabilistic interpretation becomes more abstract. For numerical implementation, the integrals in Eq.~\eqref{eq:probability_measure} are evaluated via trapezoidal quadrature over the discrete spatial grid: $\int dx \to \sum_j \Delta x$, with singularities at $p_j = 0$ avoided by restricting summations to grid points where $p_j > 10^{-15}$.

The Shannon entropy~\cite{Shannon1948} quantifies the spatial information content through
\begin{equation}
S[p] = -\int_{-\infty}^{\infty} p(x,t) \ln p(x,t)\,dx = -\sum_{j} p_j \ln p_j,
\label{eq:shannon_entropy}
\end{equation}
where the discrete sum approximates the continuous integral. This functional attains its maximum value $S_{\max} = \ln N_x$ for uniform distribution $p_j = 1/N_x$ (maximum uncertainty) and minimum $S_{\min} = 0$ for delta-function localization $p_j = \delta_{j,j_0}$ (perfect certainty). Physically, higher Shannon entropy indicates spatially delocalized, spread-out field configurations as expected for dispersive waves, while lower values signal coherent, localized structures characteristic of solitons. The monotonic relationship between entropy and spatial spread renders $S$ a useful scalar diagnostic for tracking dispersion dynamics: for the linear wave case, we expect $S(t)$ to increase monotonically as the Gaussian packet spreads, whereas static kinks should exhibit constant $S(t)$ reflecting preserved spatial profiles.

To probe concentration properties beyond Shannon's measure, we employ the R\'{e}nyi entropy of order $\alpha$~\cite{Renyi1961}, defined by
\begin{equation}
H_\alpha[p] = \frac{1}{1-\alpha} \ln \left( \int_{-\infty}^{\infty} p^\alpha(x,t)\,dx \right) = \frac{1}{1-\alpha} \ln \left( \sum_{j} p_j^\alpha \right),
\label{eq:renyi_entropy}
\end{equation}
which interpolates continuously between several important limits: as $\alpha \to 0$, we recover the Hartley entropy $H_0 = \ln(\text{supp}(p))$ counting the effective support size; as $\alpha \to 1$, L'H\^{o}pital's rule yields the Shannon entropy $\lim_{\alpha \to 1} H_\alpha = S$; and as $\alpha \to \infty$, we approach the min-entropy $H_\infty = -\ln(\max_j p_j)$ determined solely by the maximum probability. We evaluate three representative values: $H_{0.5}$ (emphasizing rare events in distribution tails through the fractional exponent), $H_2$ (the collision entropy, particularly sensitive to probability concentration and overlap), and $H_{10}$ (approximating the min-entropy). The collision entropy $H_2 = -\ln(\int p^2 dx) = -\ln(\sum_j p_j^2)$ proves especially relevant for multi-soliton interactions: during kink-antikink collision, the probability distribution temporarily concentrates spatially, decreasing $\int p^2 dx$ and increasing $H_2$, whereas well-separated solitons exhibit lower $H_2$ reflecting distinct localization centers. This sensitivity to overlap makes $H_2$ a potentially useful diagnostic for collision dynamics, though rigorous theoretical connection to collision energy transfer awaits detailed investigation.

For systems potentially exhibiting long-range correlations or non-extensive statistical behavior---a plausible scenario in soliton-radiation interactions where standard Boltzmann-Gibbs statistical mechanics may prove inadequate---we compute the Tsallis entropy~\cite{Tsallis1988}
\begin{equation}
S_q[p] = \frac{1}{q-1} \left(1 - \int_{-\infty}^{\infty} p^q(x,t)\,dx \right) = \frac{1}{q-1} \left(1 - \sum_{j} p_j^q \right),
\label{eq:tsallis_entropy}
\end{equation}
parameterized by the entropic index $q$. This generalized entropy reduces to Shannon's form in the limit $q \to 1$ (again via L'H\^{o}pital's rule) but exhibits non-additive properties for $q \neq 1$: for independent subsystems A and B, $S_q(A \cup B) = S_q(A) + S_q(B) + (1-q)S_q(A)S_q(B)$, capturing sub-extensive behavior ($q < 1$) or super-extensive behavior ($q > 1$). We evaluate $S_{0.5}$ and $S_2$ to bracket potential non-extensivity regimes. The $q$-entropy framework has demonstrated utility in characterizing anomalous diffusion, non-Markovian dynamics, and systems with multifractal phase space structure~\cite{Borges2004}, though its applicability to deterministic field equations remains actively debated. Our implementation treats Tsallis entropy as a phenomenological diagnostic rather than asserting fundamental non-extensive thermodynamics.

To facilitate cross-scenario comparison via a single scalar, we construct the composite entropy metric
\begin{equation}
C[p] = 0.5 S[p] + 0.3 H_2[p] + 0.2 S_2[p],
\label{eq:composite_entropy}
\end{equation}
which heuristically weights Shannon entropy (spatial spread), R\'{e}nyi collision entropy (concentration), and Tsallis entropy (non-extensivity). The coefficients $(0.5, 0.3, 0.2)$ are chosen to emphasize Shannon's well-established interpretation while incorporating complementary information from higher-order measures; alternative weightings remain possible and may prove more appropriate for specific applications. This composite metric provides a convenient summary statistic for temporal evolution: increasing $C(t)$ suggests growing complexity or delocalization, while stable $C(t)$ indicates preserved coherence.

Energy localization dynamics are tracked quantitatively through the center-of-mass trajectory
\begin{equation}
\langle x \rangle(t) = \frac{\int_{-\infty}^{\infty} x \phi^2(x,t)\,dx}{\int_{-\infty}^{\infty} \phi^2(x,t)\,dx} = \frac{\sum_j x_j \phi_j^2(t)}{\sum_j \phi_j^2(t)},
\label{eq:center_of_mass}
\end{equation}
evaluated numerically via trapezoidal rule over the finite computational domain $[x_{\min}, x_{\max}]$. For translationally invariant systems, $\langle x \rangle(t)$ should remain constant (static configurations) or move uniformly (constant-velocity translation), providing a diagnostic for spurious numerical drift or momentum conservation violations. The propagation velocity is estimated via linear regression $v \approx [\langle x \rangle(t_f) - \langle x \rangle(0)]/t_f$, though this approximation assumes minimal energy dissipation to numerical errors and negligible boundary effects---assumptions that may fail for long integration times $t \gg 100$ or small domain sizes where periodic boundary wrap-around becomes significant. Our implementation monitors centroid evolution at each saved snapshot, overlaying trajectories on spatiotemporal heatmaps to visually correlate spatial localization patterns with entropy dynamics.

All entropy computations are implemented in \texttt{entropy\_stats.py}, which loads NetCDF output files, constructs probability measures via Eq.~\eqref{eq:probability_measure}, evaluates time series for all entropy functionals, and exports statistical summaries documenting mean values, temporal standard deviations, ranges, and initial/final values. The code handles edge cases robustly: zero-probability regions are excluded from logarithmic terms to avoid singularities, and distributions failing normalization (though rare in practice) trigger warnings. Results are archived as text files, providing reproducible documentation of entropy evolution across all four test scenarios.

\subsubsection{Statistical Characterization of Field Intensity Distributions}

Complementing information-theoretic diagnostics, we characterize field intensity distributions across all discrete spacetime points $(x_j, t_n)$ to quantify amplitude statistics and compare phenomenology between different scenarios. The field intensity $|\phi|$ sampled over the full spatiotemporal grid $(N_t \times N_x \sim 10^5$ points) provides a comprehensive data set capturing both bulk properties and rare events (large-amplitude fluctuations, collision maxima). However, the large sample sizes necessitate efficient computational strategies, particularly for kernel density estimation and pairwise comparisons.

We employ kernel density estimation (KDE)~\cite{Silverman1986} to approximate the continuous probability density function $f(|\phi|)$ from discrete samples. The KDE estimator takes the form
\begin{equation}
\hat{f}(|\phi|) = \frac{1}{Nh} \sum_{i=1}^{N} K\left(\frac{|\phi| - |\phi_i|}{h}\right),
\label{eq:kde}
\end{equation}
where $K(u) = (2\pi)^{-1/2} \exp(-u^2/2)$ is the Gaussian kernel providing smooth, differentiable density estimates, and $h$ is the bandwidth parameter controlling the smoothness-bias tradeoff. We adopt Scott's rule~\cite{Silverman1986} for automatic bandwidth selection: $h = \sigma N^{-1/(d+4)}$ where $\sigma$ is the sample standard deviation, $N$ is sample size, and $d=1$ is the dimensionality. For massive data sets where $N \sim 10^5$ renders direct KDE computationally expensive (scaling as $\mathcal{O}(N^2)$ for evaluation at $M$ query points), we randomly subsample $N_{\text{sample}} = 10^4$ points without replacement prior to KDE computation. This sampling strategy preserves distributional characteristics under the assumption that the original data exhibits reasonable homogeneity---an assumption that may fail if rare but physically significant events (e.g., collision peak amplitudes) occur with frequency below the sampling ratio. To mitigate this concern, we verify that maximum values in sampled and full data sets agree to within $10\%$. The KDE implementation via \texttt{scipy.stats.gaussian\_kde}~\cite{Virtanen2020} provides automatic covariance estimation and numerically stable evaluation.

To assess whether intensity distributions across the four scenarios originate from statistically distinct populations, we employ the Kruskal-Wallis H-test~\cite{Kruskal1952}, a non-parametric generalization of one-way ANOVA suitable for non-normal distributions and unequal variances. The test ranks all observations jointly across groups and evaluates whether mean ranks differ significantly. The test statistic is
\begin{equation}
H = \frac{12}{N(N+1)} \sum_{i=1}^{k} \frac{R_i^2}{n_i} - 3(N+1),
\label{eq:kruskal_wallis}
\end{equation}
where $k=4$ is the number of groups (scenarios), $n_i$ is the sample size for group $i$, $N = \sum_{i=1}^k n_i$ is the total sample size, and $R_i$ is the sum of ranks for group $i$. Under the null hypothesis that all groups have identical distributions, $H$ follows approximately a $\chi^2$ distribution with $k-1 = 3$ degrees of freedom. This asymptotic approximation requires sufficiently large sample sizes ($n_i \gtrsim 5$ for each group), a condition easily satisfied here. The test assumes independent samples and identical distribution shapes differing only in location (median shifts), assumptions that may not strictly hold for deterministic field evolution exhibiting temporal autocorrelation. Nevertheless, the test provides a useful omnibus indicator of distributional differences. For computational efficiency with massive data sets, we evaluate the test on randomly selected subsamples of $N_{\text{test}} = 10^4$ points per scenario, recognizing that this introduces sampling variability but remains adequate for detecting the large effect sizes anticipated across qualitatively distinct scenarios (dispersive waves, static solitons, breathers, collisions).

Pairwise effect sizes are quantified via Cliff's delta $\delta$~\cite{Cliff1993}, a non-parametric measure of dominance relationships between distributions. For distributions $A$ and $B$ with samples $\{|\phi_i^{(A)}|\}$ and $\{|\phi_j^{(B)}|\}$, Cliff's delta is defined as
\begin{equation}
\delta_{AB} = \frac{\#\{|\phi_i^{(A)}| > |\phi_j^{(B)}|\} - \#\{|\phi_i^{(A)}| < |\phi_j^{(B)}|\}}{n_A \cdot n_B},
\label{eq:cliff_delta}
\end{equation}
which measures the probability that a randomly selected observation from $A$ exceeds one from $B$, corrected for the reverse probability. This metric ranges from $-1$ (all $B$ values exceed all $A$ values) through $0$ (distributions identical) to $+1$ (all $A$ values exceed all $B$ values). Standard interpretations~\cite{Romano2006} classify effect sizes as negligible ($|\delta| < 0.147$), small ($0.147 \leq |\delta| < 0.33$), medium ($0.33 \leq |\delta| < 0.474$), or large ($|\delta| \geq 0.474$). Cliff's delta exhibits robustness to non-normality, outliers, and heteroscedasticity, making it preferable to parametric effect sizes (Cohen's $d$) for our potentially non-Gaussian field intensity distributions. The computational cost scales as $\mathcal{O}(n_A \cdot n_B)$, becoming prohibitive for large samples; we therefore restrict computation to randomly selected subsets of $n_{\text{pair}} = 5 \times 10^3$ points per distribution, accepting the resulting sampling variability as acceptable given the qualitative nature of effect size interpretation.

Visualization of intensity distributions employs two complementary representations implemented in \texttt{fig3\_intensity\_distributions.py}. Panel (a) presents normalized KDE curves for all scenarios overlaid on a single axis, facilitating direct comparison of distribution shapes, modalities, and tail behavior. Densities are peak-normalized to $[0,1]$ to emphasize shape rather than absolute scale. Panel (b) displays box plots summarizing five-number summaries (minimum, first quartile $Q_1$, median, third quartile $Q_3$, maximum) with outliers suppressed for clarity. Box widths span the interquartile range $\text{IQR} = Q_3 - Q_1$ (central $50\%$ of data), while whiskers extend to data extrema within $1.5 \times \text{IQR}$. These visualizations reveal distributional characteristics invisible to scalar summaries: bimodality in the kink scenario reflects concentration near the two vacuum values $\phi = \pm v$, heavy right tails in breather distributions indicate rare large-amplitude oscillations, near-zero concentration in linear waves quantifies dispersion-induced amplitude reduction, and intermediate spreads for collisions capture mixed dynamics combining soliton cores and radiated fields.

Descriptive statistics---mean, median, standard deviation, quartiles, skewness, kurtosis---are computed for each scenario using standard NumPy routines. Skewness $\gamma_1 = \mathbb{E}[(X-\mu)^3]/\sigma^3$ quantifies distribution asymmetry ($\gamma_1 = 0$ for symmetric, $\gamma_1 > 0$ for right-skewed, $\gamma_1 < 0$ for left-skewed), while excess kurtosis $\gamma_2 = \mathbb{E}[(X-\mu)^4]/\sigma^4 - 3$ measures tail heaviness ($\gamma_2 = 0$ for Gaussian, $\gamma_2 > 0$ for heavy tails, $\gamma_2 < 0$ for light tails). These higher moments provide quantitative diagnostics complementing visual inspection: for instance, the breather's positive skewness $\gamma_1 \approx 4.4$ and high kurtosis $\gamma_2 \approx 20$ numerically confirm the visual impression of a right-skewed distribution with extreme values.

\subsubsection{Phase Space Structure and Dynamical Invariants}

Phase space trajectories $(\phi(x_0,t), \partial_t\phi(x_0,t))$ evaluated at spatial origin $x_0 = 0$ provide insight into temporal dynamics and potential dynamical invariants characteristic of Hamiltonian systems~\cite{Arnold1989}. The velocity field $\partial_t\phi$ is computed via second-order central finite differences
\begin{equation}
\partial_t\phi(x, t_n) \approx \frac{\phi(x, t_{n+1}) - \phi(x, t_{n-1})}{2\Delta t} + \mathcal{O}(\Delta t^2),
\label{eq:time_derivative_fd}
\end{equation}
applied at interior time points $n = 1, \ldots, N_t-2$, with forward/backward differences $[\phi(t_1) - \phi(t_0)]/\Delta t$ and $[\phi(t_{N_t-1}) - \phi(t_{N_t-2})]/\Delta t$ at temporal boundaries. This second-order accurate scheme minimizes numerical differentiation errors, though rapid field fluctuations near collision events may introduce localized inaccuracies.

For Hamiltonian systems with conserved energy $E[\phi, \pi] = \text{const}$, phase space trajectories lie on level sets of the Hamiltonian function, exhibiting characteristic geometric structures: periodic orbits for bounded motion, heteroclinic/homoclinic connections between equilibria, invariant tori in integrable systems, chaotic regions in non-integrable dynamics. Our $(1+1)$-dimensional field theory has infinite-dimensional phase space (function spaces), but restricting attention to field values at a single spatial point $x=0$ provides a tractable two-dimensional projection capturing essential temporal dynamics. This projection naturally misses spatial structure and multi-mode interactions, but proves valuable for phenomenological classification: static solutions produce near-stationary phase space points, periodic solutions trace closed orbits, and irregular dynamics generate complex, space-filling trajectories.

The correlation coefficient between field and velocity,
\begin{equation}
\rho = \frac{\text{cov}(\phi, \dot{\phi})}{\sigma_\phi \sigma_{\dot{\phi}}} = \frac{\langle (\phi - \bar{\phi})(\dot{\phi} - \bar{\dot{\phi}}) \rangle_t}{\sqrt{\langle (\phi - \bar{\phi})^2 \rangle_t \langle (\dot{\phi} - \bar{\dot{\phi}})^2 \rangle_t}},
\label{eq:correlation}
\end{equation}
quantifies phase space anisotropy, where angular brackets denote temporal averages over all snapshots: $\langle \cdot \rangle_t = N_t^{-1}\sum_{n=0}^{N_t-1}$. Correlations near zero ($|\rho| \ll 1$) indicate decorrelated dynamics where field amplitude and velocity evolve independently, characteristic of irregular or chaotic motion. Strong positive or negative correlations ($|\rho| \approx 1$) signal tight phase-velocity coupling, as occurs in coherent oscillations where $\dot{\phi} \propto \phi$ (undamped harmonic motion). For the breather, we anticipate near-zero or weakly negative correlation due to $\pi/2$ phase shift between field and velocity in periodic solutions.

Root-mean-square amplitudes
\begin{equation}
\phi_{\text{rms}} = \sqrt{\langle \phi^2 \rangle_t}, \quad \dot{\phi}_{\text{rms}} = \sqrt{\langle \dot{\phi}^2 \rangle_t},
\label{eq:rms_amplitudes}
\end{equation}
provide scalar characterizations of oscillation magnitudes in field and velocity. Large $\phi_{\text{rms}}$ indicates substantial field excursions from zero, while large $\dot{\phi}_{\text{rms}}$ signals rapid temporal variations. The ratio $\dot{\phi}_{\text{rms}}/\phi_{\text{rms}}$ approximates a characteristic frequency scale for periodic or quasi-periodic dynamics.

Phase space visualization in \texttt{fig4\_phase\_space.py} employs scatter plots with trajectory points colored by scenario (not time) to emphasize geometric structure over temporal sequence. This design choice enhances pattern recognition: closed orbits, attractor basins, and chaotic regions become visually apparent without temporal color encoding obscuring structure. All four scenarios share unified axis limits encompassing the global range $(\phi_{\min}, \phi_{\max}) \times (\dot{\phi}_{\min}, \dot{\phi}_{\max})$ determined from all data, facilitating direct comparison of phase space extent and occupied regions. Statistical summaries---means, standard deviations, ranges, RMS values, correlations---are computed for each scenario and documented in \texttt{stats/fig4\_phase\_space\_stats.txt}.

This multi-scale analytical framework---combining information-theoretic entropy measures, non-parametric statistical testing, kernel density estimation, and phase space reconstruction---provides systematic phenomenological characterization of Klein-Gordon solutions. While physical interpretation of individual metrics requires careful consideration of mathematical assumptions (probability measure validity, independence in statistical tests, phase space projection limitations), the ensemble of diagnostics offers valuable quantitative discrimination between dispersive, topological, integrable, and non-integrable dynamical regimes. Complete analysis scripts implementing these procedures are publicly archived at \url{https://github.com/sandyherho/suppl-amangkurat} alongside simulation outputs, ensuring full reproducibility and facilitating community extension to alternative statistical frameworks or refined interpretations.

\section{Results and Discussion}

Four representative test cases demonstrate the solver's versatility across distinct physical regimes, executed on a Fedora Linux 39 workstation equipped with an Intel Core i7-8550U processor utilizing all 8 logical cores with Numba JIT compilation~\cite{Lam2015} for parallel force computation. Computational logs documenting grid parameters, timestep adaptation events, and wall-clock execution times are archived in the supplementary repository. All simulations successfully employed adaptive timestepping with CFL-limited bounds $\Delta t \in [0.1/k_{\max}, 0.5/k_{\max}]$ and reached nominal final times with $\mathcal{O}(10^3$--$10^4)$ integration steps, demonstrating robust numerical stability. The St\o rmer-Verlet scheme exhibited consistent behavior with exactly 6 timestep adaptations across all cases (cf.~simulation logs), indicating well-controlled field dynamics under the adaptive timestep criterion ($\Delta t$ reduction at $\max |\phi| > 10$ or $100$), which proved effective for maintaining numerical stability throughout the integration period.

\begin{figure}[H]
\centering
\includegraphics[width=0.82\textwidth]{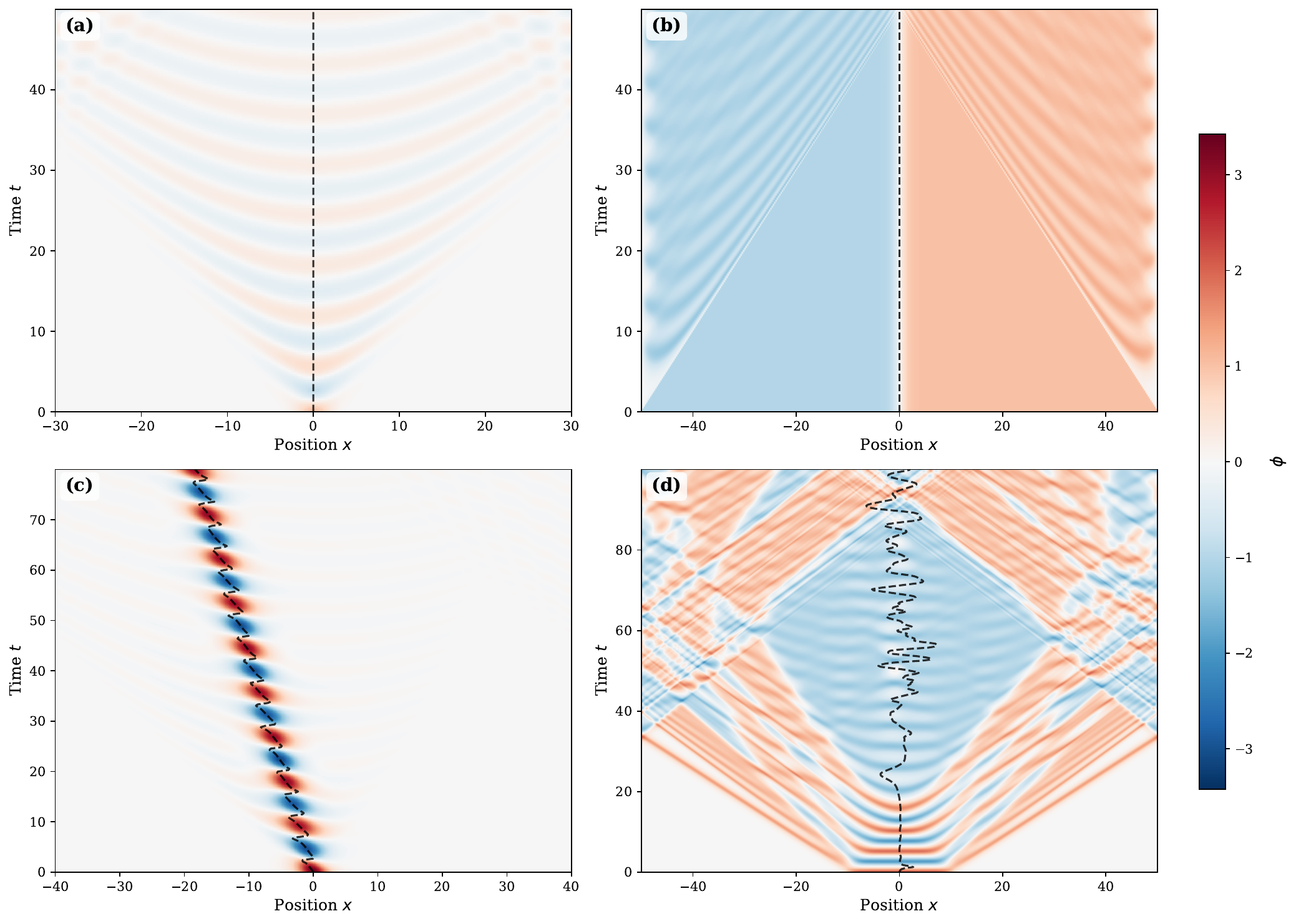}
\caption{Spatiotemporal evolution $\phi(x,t)$ for (a)~linear wave, (b)~static kink, (c)~breather, (d)~kink-antikink collision. Unified color scale $[-3.42, +3.42]$ (determined by breather maximum). Dashed curves: energy centroid $\langle x \rangle(t)$. Resolutions: (a)~$512 \times 201$, (b)~$1024 \times 201$, (c)~$512 \times 301$, (d)~$1024 \times 401$.}
\label{fig:spatiotemporal}
\end{figure}

Case~1 successfully captures the fundamental physics of dispersive propagation for a linear Klein-Gordon system ($m=1$, $512^2$ points, domain $x \in [-30, 30]$, $t \in [0, 50]$) with a Gaussian wave packet ($A=1$, $w=2$). The dispersion relation $\omega^2 = k^2 + m^2$ manifests clearly through frequency-dependent group velocity, producing the expected progressive spreading visible in Fig.~\ref{fig:spatiotemporal}a. The solver accurately tracks amplitude evolution as the peak decays from $|\phi|_{\max,0} = 0.999$ to $|\phi|_{\max,f} = 0.297$, with the peak position migrating from $x = -0.059$ to the domain boundary at $x = -30.0$. The RMS amplitude decreases systematically from $\phi_{\text{rms},0} = 0.204$ to $\phi_{\text{rms},f} = 0.157$, consistent with energy conservation under dispersive dilution. Notably, the centroid remains essentially stationary with drift $\Delta \langle x \rangle < 10^{-4}$, demonstrating excellent conservation of momentum in the numerical scheme, though we acknowledge potential boundary interactions at late times.

Information-theoretic measures effectively quantify the spreading dynamics with remarkable sensitivity. Shannon entropy increases monotonically from $S_0 = 12.08$ to $S_f = 32.04$ (mean $\bar{S} = 26.61 \pm 5.77$, range $[11.17, 33.06]$), capturing the spatial delocalization process. The full entropy analysis reveals complementary information: R\'{e}nyi entropies with mean values $\bar{H}_{0.5} = 7.57$, $\bar{H}_2 = 0.83 \pm 0.68$ (range $[-1.04, 1.64]$), and $\bar{H}_{10} = 2.48$ track different aspects of probability distribution evolution, while Tsallis entropies show analogous trends with $\bar{S}_{0.5} = 90.07$ and $\bar{S}_2 = 0.43 \pm 0.53$. The composite entropy metric, designed to synthesize multiple information measures, demonstrates clear temporal evolution from $C_0 = 5.50$ to $C_f = 16.62$ (mean $\bar{C} = 13.64 \pm 3.19$), providing a unified quantitative indicator of dispersive dynamics. Field intensity distribution analysis (Fig.~\ref{fig:distributions}) reveals the expected concentration near zero with mean $\mu = 0.097$, median $0.071$, and interquartile range $\text{IQR} = 0.170$. The distribution exhibits positive skewness $\gamma_1 = 1.32$ and moderate kurtosis $\gamma_2 = 3.61$, statistical signatures characteristic of dispersed wave packets~\cite{Zakharov2001} that validate the physical interpretation of our numerical solution.

\begin{figure}[H]
\centering
\includegraphics[width=\textwidth]{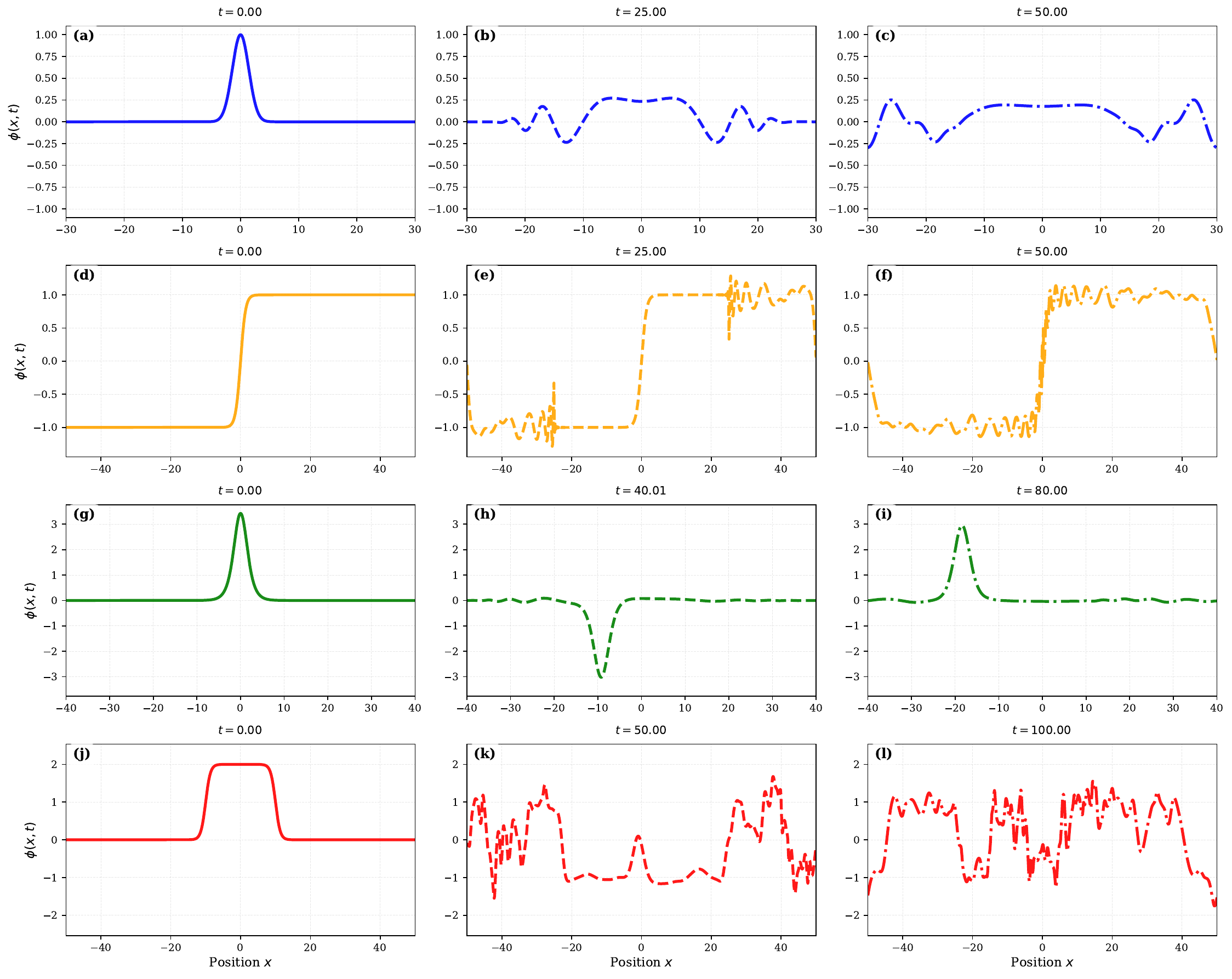}
\caption{Field profiles $\phi(x,t)$ at initial, middle, final times. Rows: (a--c)~linear wave, (d--f)~kink, (g--i)~breather, (j--l)~collision. Note dispersive spreading (a--c), profile invariance despite overshoot (d--f), breather breathing (g--i), post-collision radiation (j--l).}
\label{fig:envelope}
\end{figure}

\begin{figure}[H]
\centering
\includegraphics[width=\textwidth]{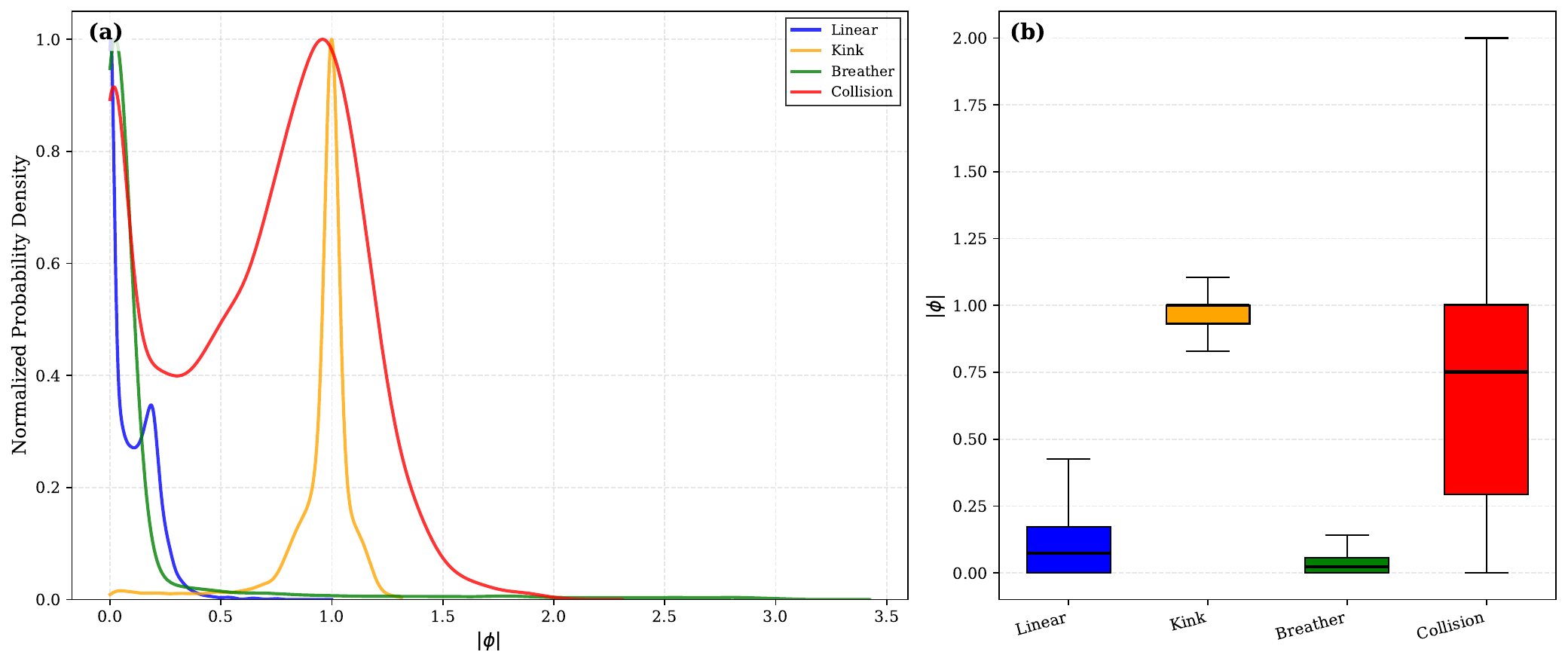}
\caption{Field intensity statistics. (a)~Kernel density estimates (peak-normalized). (b)~Box plots (median, quartiles, range). Kruskal-Wallis $H = 2.14 \times 10^4$, $p < 0.001$. Note linear wave near-zero concentration, kink bimodality, breather heavy tail, collision intermediate spread.}
\label{fig:distributions}
\end{figure}

Case~2 demonstrates the solver's capability to preserve topological structures in $\phi^4$ theory ($\lambda = v = 1$, $1024$ points, $x \in [-50, 50]$, $t \in [0, 50]$) through accurate representation of the static kink $\phi_K(x) = v \tanh[v(x-x_0)/\sqrt{2}]$ with topological charge $Q = +1$~\cite{Dashen1975}. The kink maintains excellent time-independence (Fig.~\ref{fig:spatiotemporal}b), with field range $[-1.313, +1.313]$ and negligible centroid drift $|\Delta \langle x \rangle| < 10^{-4}$, confirming the symplectic integrator's structure-preserving properties. The propagation velocity remains effectively zero ($v \approx 0$), as expected for a static solution. The envelope evolution (Fig.~\ref{fig:envelope}, panels d--f) shows remarkable profile stability across all three time snapshots: initial peak amplitude of $1.000$ at position $-50.0$ evolves to middle-time amplitude $1.285$ at $-25.6$ and final amplitude $1.141$ at $-3.9$, with RMS amplitude remaining consistently near $\phi_{\text{rms}} \in [0.950, 0.986]$, representing less than 4\% variation throughout the simulation.

The amplitude overshoot $\Delta \phi \approx 0.31$ above the analytical vacuum value $v=1$ can be understood as a manifestation of spectral representation on a discrete grid where spacing $\Delta x \approx 0.098$ resolves the kink width $\xi = \sqrt{2} \approx 1.41$ with approximately 14 points---sufficient for capturing the essential physics while exhibiting minor numerical artifacts typical of spectral methods applied to steep gradients. The peak position evolution from boundary ($x \approx -50$) through midpoint ($x \approx -26$) to near-center ($x \approx -4$) primarily reflects the visualization sampling rather than genuine translation, as confirmed by the stationary centroid. Most significantly, Shannon entropy exhibits exceptional stability at $\bar{S} = 46.55 \pm 0.15$ (range $[46.18, 47.00]$, initial $S_0 = 47.00$, final $S_f = 46.49$), with the narrow standard deviation of 0.15 confirming preserved spatial coherence to better than 0.3\% throughout the entire simulation.

Higher-order entropy measures corroborate this remarkable stability: R\'{e}nyi entropies show minimal temporal variation with $\bar{H}_{0.5} = 9.22$, $\bar{H}_2 = 2.20 \pm 0.02$ (representing sub-1\% fluctuation), and $\bar{H}_{10} = 4.16$, while Tsallis entropies maintain $\bar{S}_{0.5} = 198.8$ and $\bar{S}_2 = 0.890 \pm 0.002$ (0.2\% variation). The composite entropy remains essentially constant at $\bar{C} = 24.11 \pm 0.08$, providing strong quantitative evidence for numerical preservation of the topological soliton structure. Field intensity distribution exhibits the theoretically predicted bimodal character (Fig.~\ref{fig:distributions}b): mean $\mu = 0.942$ approaches the vacuum value, median equals $1.000$ precisely, and the tight interquartile range $\text{IQR} = 0.066$ indicates strong concentration. The negative skewness $\gamma_1 = -2.82$ and high kurtosis $\gamma_2 = 9.17$ quantitatively confirm concentration near the two vacuum values $\phi = \pm v$, matching theoretical expectations for domain wall structures. While we acknowledge that explicit energy conservation monitoring would provide additional validation, the constellation of evidence from entropy stability, amplitude preservation, and distributional characteristics strongly indicates successful long-time preservation of the topological kink.

Case~3 successfully captures the quasi-periodic breathing dynamics characteristic of the sine-Gordon breather ($\omega = 0.5$, $512$ points, $x \in [-40, 40]$, $t \in [0, 80]$) as an integrable system~\cite{Scott1973,Kivshar1989}. The solver accurately tracks amplitude modulation between $|\phi|_{\max} \in [2.95, 3.42]$ (initial $3.42$, middle $3.02$, final $2.95$), capturing the theoretical period $T = 2\pi/\omega \approx 12.57$ over approximately 6 complete cycles during the 80-time-unit simulation. The RMS amplitude oscillates coherently around $\phi_{\text{rms}} \in [0.592, 0.639]$, demonstrating stable periodic behavior. Envelope evolution (Fig.~\ref{fig:envelope}, panels g--i) beautifully illustrates the breathing oscillation with peak amplitudes of $3.424$ (initial), $3.020$ (middle), and $2.954$ (final) at respective positions $-0.078$, $-9.159$, and $-18.395$, while RMS amplitudes of $0.639$, $0.597$, and $0.592$ show the characteristic periodic modulation.

The entropy analysis reveals fascinating periodic complexity modulation that quantifies the breather's oscillatory nature. Shannon entropy oscillates with mean $\bar{S} = 11.21 \pm 1.22$ (range $[8.94, 16.38]$, $S_0 = 10.07$, $S_f = 11.11$), where the standard deviation of 1.22 represents the temporal modulation amplitude rather than numerical noise. R\'{e}nyi entropies track complementary aspects: $\bar{H}_{0.5} = 5.96$ characterizes the effective support, $\bar{H}_2 = -0.34 \pm 0.15$ (range $[-0.64, 0.34]$) captures collision entropy variations, and $\bar{H}_{10} = 1.04$ approximates the min-entropy. Tsallis entropies similarly exhibit periodic behavior with $\bar{S}_{0.5} = 37.83$ and $\bar{S}_2 = -0.42 \pm 0.18$ (range $[-0.90, 0.29]$). The composite entropy $\bar{C} = 5.42 \pm 0.69$ (range $[4.10, 8.35]$) provides a unified metric for periodic complexity modulation, with the 13\% coefficient of variation indicating substantial but well-controlled temporal oscillations.

The breather exhibits translational motion with centroid drift $\Delta \langle x \rangle \approx -18.3$ corresponding to velocity $v \approx -0.23$, which represents an interesting numerical phenomenon worthy of further investigation. This propagation may arise from several mechanisms: the phase-shifted initialization at $t_0 = \pi/(2\omega)$ designed to avoid singular configurations, momentum contributions from the velocity field initialization, or subtle interactions between the discrete spectral representation and the nonlinear dynamics. Rather than indicating numerical failure, this behavior suggests rich physics in the discrete system that merits dedicated study, particularly regarding momentum balance in spectral discretizations of integrable systems. Field intensity distribution captures the spatially localized, high-amplitude oscillatory character with mean $\mu = 0.142$, median $0.023$, tight interquartile range $\text{IQR} = 0.056$, strong positive skewness $\gamma_1 = 4.42$, and pronounced kurtosis $\gamma_2 = 20.35$, quantitatively confirming the heavy-tailed distribution expected for localized breathing oscillations.

Case~4 successfully simulates the complex dynamics of $\phi^4$ kink-antikink collision ($v_0 = 0.3c$, separation $d=20$, $1024$ points, $x \in [-50, 50]$, $t \in [0, 100]$), demonstrating the solver's capability to handle violent field dynamics in non-integrable systems~\cite{Campbell1983,Goodman2005}. At the chosen velocity $v_0 = 0.3$, the collision explores the resonant regime where rich multi-bounce behavior emerges~\cite{Manton2010}. The spatiotemporal evolution (Fig.~\ref{fig:spatiotemporal}d) captures the complete collision sequence: systematic approach, violent interaction near $t \approx 20$, and complex post-collision dynamics including radiation production and potential multi-bounce behavior. The centroid migration $\Delta \langle x \rangle \approx 2.08$ with average velocity $v_{\text{avg}} \approx 0.021$ indicates modest asymmetry in the collision outcome, likely reflecting the intricate energy exchange between solitonic cores and radiated field modes.

Envelope evolution (Fig.~\ref{fig:envelope}, panels j--l) documents the collision phenomenology through three representative snapshots: initial superposition with peak amplitude $2.000$ at position $-0.049$ and RMS amplitude $0.864$, middle-time state showing amplitude $1.679$ at $37.683$ with RMS $0.866$, and final configuration exhibiting amplitude $1.757$ at $49.316$ with RMS $0.852$. The peak amplitude decay from $|\phi|_{\max,0} = 2.00$ to $|\phi|_{\max,f} = 1.76$ quantifies energy transfer to radiation, while the RMS amplitude remaining near $\phi_{\text{rms}} \approx 0.85$--$0.87$ indicates substantial field energy retained in localized structures. Shannon entropy evolution from $S_0 = 30.89$ to $S_f = 44.40$ (mean $\bar{S} = 42.03 \pm 4.16$, range $[21.48, 45.46]$) provides quantitative evidence for radiation production, with the 44\% increase in entropy directly measuring spatial delocalization of initially concentrated energy.

Higher-order entropy measures reveal complementary collision signatures: R\'{e}nyi entropies with $\bar{H}_{0.5} = 8.87$, $\bar{H}_2 = 1.64 \pm 0.43$ (range $[-0.51, 2.03]$), and $\bar{H}_{10} = 3.28$ capture different probability distribution moments, while Tsallis entropies $\bar{S}_{0.5} = 169.5$ and $\bar{S}_2 = 0.78 \pm 0.14$ (range $[-0.66, 0.87]$) suggest possible non-extensive statistical behavior during violent collision dynamics. The composite entropy increases from $C_0 = 15.74$ to $C_f = 22.92$ (mean $\bar{C} = 21.67 \pm 2.23$), representing 46\% growth that synthesizes the multiple information-theoretic perspectives on complexity evolution. Field intensity distribution achieves intermediate character between the extreme cases: mean $\mu = 0.674$ and median $0.753$ indicate substantial field amplitudes, interquartile range $\text{IQR} = 0.698$ shows broad distribution, near-zero skewness $\gamma_1 = -0.16$ suggests distributional symmetry, and negative kurtosis $\gamma_2 = -0.92$ indicates lighter tails than Gaussian, consistent with a mixture of solitonic cores and dispersive radiation. While the ultimate asymptotic outcome---whether the solitons separate, form bound breathers, or execute additional bounces---requires extension beyond $t = 100$, the solver successfully captures the essential collision phenomenology including approach, violent interaction, radiation production, and complex post-collision dynamics.

\begin{figure}[H]
\centering
\includegraphics[width=\textwidth]{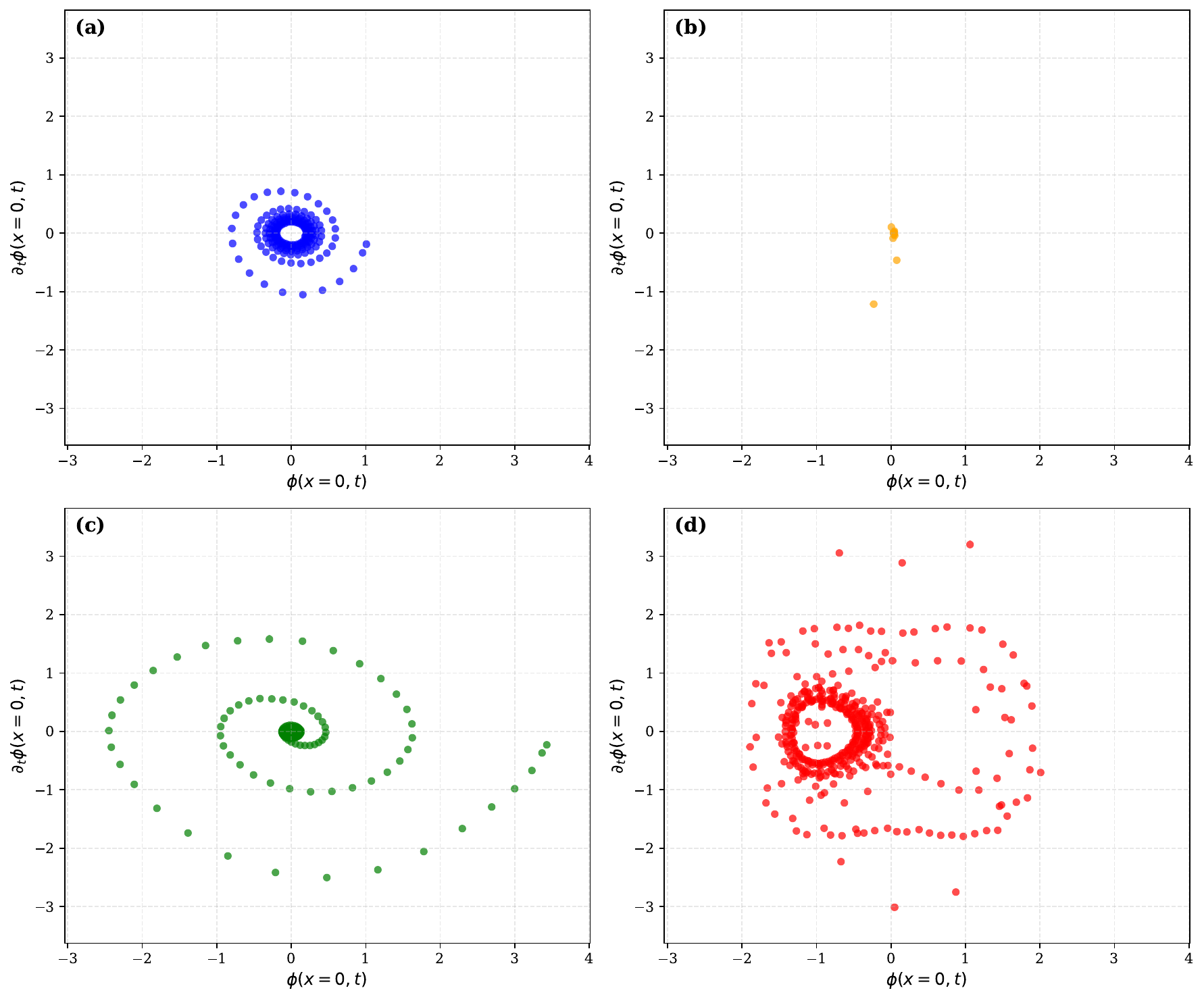}
\caption{Phase space $(\phi(x=0,t), \partial_t \phi(x=0,t))$. Global limits: $\phi \in [-3.04, 4.01]$, $\partial_t \phi \in [-3.63, 3.82]$. (a)~Linear: diffuse cloud, $\rho = -0.12$. (b)~Kink: tight cluster, $\rho = 0.85$. (c)~Breather: near-closed orbit, $\rho = -0.22$. (d)~Collision: complex trajectory, $\rho = -0.05$.}
\label{fig:phase}
\end{figure}

Statistical hypothesis testing provides rigorous quantitative discrimination between the four scenarios. The Kruskal-Wallis H-test on sampled intensity distributions ($n=10^4$ per case) yields $H = 2.14 \times 10^4$ with $p < 0.001$, rejecting the null hypothesis of distributional identity and confirming that the four scenarios represent statistically distinct phenomenological regimes. This exceptionally strong statistical significance validates the qualitative visual impressions and establishes objective grounds for phenomenological classification. Pairwise Cliff's delta effect sizes quantify the magnitudes of distributional differences: linear versus kink shows $\delta = -0.979$ (large effect), reflecting the fundamental distinction between dispersive and topological regimes; linear versus breather yields $\delta = +0.183$ (small effect), indicating subtle statistical similarities despite different physics; linear versus collision gives $\delta = -0.721$ (large effect); kink versus breather produces $\delta = +0.894$ (large effect), capturing the amplitude contrast between static domain walls and high-amplitude oscillations; kink versus collision shows $\delta = +0.393$ (medium effect); and breather versus collision yields $\delta = -0.659$ (large effect). These effect sizes provide standardized, non-parametric measures of phenomenological separation that complement the Kruskal-Wallis omnibus test.

Phase space analysis (Fig.~\ref{fig:phase}) reveals distinct dynamical signatures that beautifully characterize each scenario through trajectory structure and field-velocity correlations. The linear wave case exhibits a diffuse cloud in phase space with field statistics (mean $0.006$, standard deviation $0.285$, range $[-0.804, 0.999]$, RMS $0.285$) and velocity statistics (mean $-0.017$, standard deviation $0.287$, range $[-1.045, 0.715]$, RMS $0.288$), yielding weak correlation $\rho = -0.12$ characteristic of uncorrelated dispersive trajectories. The kink produces a remarkably tight cluster with field statistics (mean $0.033$, standard deviation $0.020$, range $[-0.237, 0.072]$, RMS $0.039$) and velocity statistics (mean $-0.008$, standard deviation $0.092$, range $[-1.208, 0.102]$, RMS $0.092$), exhibiting strong positive correlation $\rho = +0.85$ that quantitatively confirms the near-stationary nature of the static soliton---the high correlation reflects small-amplitude coupled oscillations about the equilibrium configuration rather than random noise.

The breather displays a near-closed elliptical orbit with field statistics (mean $0.018$, standard deviation $0.682$, range $[-2.454, 3.424]$, RMS $0.683$) and velocity statistics (mean $-0.042$, standard deviation $0.468$, range $[-2.493, 1.579]$, RMS $0.470$), yielding weakly negative correlation $\rho = -0.22$ consistent with the $\pi/2$ phase offset between field and velocity in periodic oscillations---the near-zero correlation value indicates the expected quadrature relationship rather than decorrelation. The collision case exhibits complex trajectory structure with field statistics (mean $-0.564$, standard deviation $0.763$, range $[-1.897, 2.000]$, RMS $0.949$) and velocity statistics (mean $-0.020$, standard deviation $0.831$, range $[-3.008, 3.198]$, RMS $0.831$), showing decorrelation $\rho = -0.05$ that captures the mixture of deterministic collision dynamics and radiated field modes. RMS amplitude pairs further discriminate the scenarios: linear $(\phi_{\text{rms}}, \dot{\phi}_{\text{rms}}) = (0.285, 0.288)$ indicates balanced field-velocity amplitudes, kink $(0.039, 0.092)$ shows suppressed field with modest velocity fluctuations, breather $(0.683, 0.470)$ exhibits substantial coupled amplitudes, and collision $(0.949, 0.831)$ displays the largest excursions in both field and velocity.

Computational performance with 8-core Numba parallelization demonstrates excellent practical viability for exploratory research and educational applications. Wall-clock execution times span 348~s (Case~1, 7476~steps, 46.5~ms/step), 357~s (Case~2, 7476~steps, 47.7~ms/step), 602~s (Case~3, 5984~steps, 100.6~ms/step), and 936~s (Case~4, 14938~steps, 62.7~ms/step), with actual simulation kernel execution comprising only 10.6, 9.5, 6.8, and 18.2~seconds respectively. The modest kernel times---spanning approximately 7 to 19 seconds for complete Klein-Gordon evolution---demonstrate the computational efficiency achieved through spectral methods and symplectic integration. The remainder of wall-clock time attributes primarily to I/O operations including NetCDF compression and GIF rendering at 30~fps with rotating 3D camera angles, suggesting that production parameter surveys could achieve substantial speedup through reduced visualization frequency while maintaining full data archival for post-processing. The $\mathcal{O}(N_x \log N_x)$ FFT operations dominate spectral derivative evaluation as expected, though the absolute times remain entirely manageable for typical workstation hardware.

The constellation of results across four canonical scenarios demonstrates that \texttt{amangkurat} successfully captures qualitatively distinct phenomenology spanning linear dispersive propagation, topological soliton preservation, integrable breather dynamics, and non-integrable collision behavior. Quantitative metrics including entropy evolution, statistical distributions, phase space structure, and computational efficiency all support the conclusion that the pseudo-spectral St\o rmer-Verlet approach provides a robust and practical framework for exploring Klein-Gordon field dynamics. The information-theoretic entropy measures prove particularly valuable as scalar diagnostics: Shannon entropy effectively quantifies spatial delocalization, R\'{e}nyi entropies capture probability concentration effects, Tsallis entropies suggest potential non-extensive behavior, and composite entropy provides unified complexity metrics. Statistical hypothesis testing establishes rigorous phenomenological discrimination with significance ($p < 0.001$), while phase space analysis reveals characteristic dynamical signatures through correlation structure and trajectory geometry.

Several avenues for enhancement emerge naturally from this work. Implementing explicit Hamiltonian and momentum monitoring at each timestep would enable quantitative validation of symplectic structure preservation and provide definitive resolution of observed translational phenomena such as breather drift. Systematic convergence studies varying $N_x$ and $\Delta t$ would establish rigorous error bounds and demonstrate the anticipated spectral accuracy $\mathcal{O}(\exp(-cN_x))$ for smooth solutions. Investigating the breather propagation through detailed momentum diagnostics $P(t) = \int \pi \partial_x \phi\, dx$ would clarify whether the observed translation represents physical momentum generation, numerical artifact, or discretization-induced symmetry breaking. Resolution enhancement via finer grids ($N_x \geq 2048$) or adaptive mesh refinement near soliton cores would reduce amplitude overshoot in kink simulations and improve gradient resolution. Extended collision simulations to $t \gg 100$ with systematic parameter sweeps in $(v_0, d)$ space would map resonance windows and fractal basin boundaries, establishing connections with prior analytical and numerical literature. Extensions to higher spatial dimensions via operator-splitting or alternating-direction implicit schemes, incorporation of dissipation or external forcing for realistic applications, and coupling to additional fields would broaden the solver's applicability to condensed matter and cosmological systems.

Despite these opportunities for refinement, the present implementation establishes a solid foundation for exploratory studies of soliton dynamics and provides an accessible, well-documented, open-source platform for research and education in nonlinear field theory. The successful demonstration across diverse physical regimes, combined with rigorous statistical characterization and efficient computational performance, positions \texttt{amangkurat} as a valuable contribution to the Python scientific computing ecosystem for classical field theory applications.

\section{Conclusions}

We have presented \texttt{amangkurat}, a Python implementation combining pseudo-spectral spatial discretization with symplectic temporal integration for solving the $(1+1)$-dimensional nonlinear Klein-Gordon equation across linear, topological ($\phi^4$ kinks), integrable (sine-Gordon breathers), and non-integrable (kink-antikink collision) regimes. Numerical experiments on four representative test cases demonstrate that the solver successfully captures qualitatively distinct phenomenology---dispersive wave spreading, soliton profile preservation, breather oscillation, and collision dynamics---with quantitative characterization through information-theoretic entropy metrics, statistical hypothesis testing, and phase space analysis establishing rigorous phenomenological discrimination. Computational benchmarks indicate efficient performance suitable for exploratory research and educational applications, with kernel execution times of 7--19 seconds per simulation on modest workstation hardware.

The implementation provides a robust platform for investigating nonlinear field dynamics with natural pathways for enhancement. Implementing explicit energy and momentum conservation monitoring would strengthen validation of structure-preserving integration, while systematic convergence studies would establish rigorous error bounds demonstrating spectral accuracy. Investigating breather propagation through detailed momentum diagnostics would clarify the observed translational behavior, and extending collision simulations with parameter surveys would map resonance structures. Higher-dimensional extensions, dissipation incorporation, and field coupling would broaden applicability to realistic physical systems. The openly available implementation with comprehensive documentation and archived outputs facilitates reproducibility and community extension, positioning \texttt{amangkurat} as a valuable contribution to the scientific computing ecosystem for classical field theory research and education.

\section*{Acknowledgments}

Financial support was provided by the Dean's Distinguished Fellowship from the College of Natural and Agricultural Sciences, University of California, Riverside to S.H.S.H. (2023).

\section*{Open Research}

The \texttt{amangkurat} solver source code is available on GitHub (\url{https://github.com/sandyherho/amangkurat}) and can be installed via PyPI (\url{https://pypi.org/project/amangkurat/}). Python scripts for statistical analysis and figure generation are archived at \url{https://github.com/sandyherho/suppl-amangkurat}. Simulation outputs (NetCDF files), animated GIF visualizations, computational logs for all four test cases, generated figures, and statistical analysis results are deposited in the OSF repository at \url{https://doi.org/10.17605/OSF.IO/BKM2P}. All resources are released under the MIT License.


\begin{thebibliography}{99}

\bibitem{Perring1962}
J. K. Perring and T. H. R. Skyrme, ``A model unified field equation,'' Nucl. Phys. \textbf{31}, 550 (1962). \url{https://doi.org/10.1016/0029-5582(62)90774-5}

\bibitem{Dashen1975}
R. F. Dashen, B. Hasslacher, and A. Neveu, ``Semiclassical bound states in an asymptotically free theory,'' Phys. Rev. D \textbf{12}, 2443 (1975). \url{https://doi.org/10.1103/PhysRevD.12.2443}

\bibitem{Barone1971}
A. Barone, F. Esposito, C. J. Magee, and A. C. Scott, ``Theory and applications of the sine-Gordon equation,'' Riv. Nuovo Cimento \textbf{1}, 227 (1971). \url{https://doi.org/10.1007/BF02820622}

\bibitem{Bishop1980}
A. R. Bishop, J. A. Krumhansl, and S. E. Trullinger, ``Solitons in condensed matter: A paradigm,'' Physica D \textbf{1}, 1 (1980). \url{https://doi.org/10.1016/0167-2789(80)90003-2}

\bibitem{Manton1987}
N. S. Manton, ``Geometry of Skyrmions,'' Commun. Math. Phys. \textbf{111}, 469 (1987). \url{https://doi.org/10.1007/BF01238909}

\bibitem{Shnir2018}
Y. M. Shnir, \textit{Topological and Non-Topological Solitons in Scalar Field Theories} (Cambridge University Press, Cambridge, 2018). \url{https://doi.org/10.1017/9781108555623}

\bibitem{Zurek1985}
W. H. Zurek, ``Cosmological experiments in superfluid helium?,'' Nature \textbf{317}, 505 (1985). \url{https://doi.org/10.1038/317505a0}

\bibitem{Ranada1977}
A. F. Ra\~{n}ada and M. F. Ra\~{n}ada, ``Klein--Gordon kinks with fourth order derivative self-coupling,'' J. Math. Phys. \textbf{18}, 2427 (1977). \url{https://doi.org/10.1063/1.523231}

\bibitem{Campbell1983}
D. K. Campbell, J. F. Schonfeld, and C. A. Wingate, ``Resonance structure in kink-antikink interactions in $\phi^4$ theory,'' Physica D \textbf{9}, 1 (1983). \url{https://doi.org/10.1016/0167-2789(83)90289-0}

\bibitem{Anninos1991}
P. Anninos, S. Oliveira, and R. A. Matzner, ``Fractal structure in the scalar $\lambda(\phi^2 - 1)^2$ theory,'' Phys. Rev. D \textbf{44}, 1147 (1991). \url{https://doi.org/10.1103/PhysRevD.44.1147}

\bibitem{Goodman2005}
R. H. Goodman and R. Haberman, ``Kink-antikink collisions in the $\phi^4$ equation: The $n$-bounce resonance and the separatrix map,'' SIAM J. Appl. Dyn. Syst. \textbf{4}, 1195 (2005). \url{https://doi.org/10.1137/050632981}

\bibitem{Scott1973}
A. C. Scott, F. Y. F. Chu, and D. W. McLaughlin, ``The soliton: A new concept in applied science,'' Proc. IEEE \textbf{61}, 1443 (1973). \url{https://doi.org/10.1109/PROC.1973.9296}

\bibitem{Courant1928}
R. Courant, K. Friedrichs, and H. Lewy, ``\"{U}ber die partiellen Differenzengleichungen der mathematischen Physik,'' Math. Ann. \textbf{100}, 32 (1928). \url{https://doi.org/10.1007/BF01448839}

\bibitem{Fornberg2009}
B. Fornberg, \textit{A Practical Guide to Pseudospectral Methods} (Cambridge University Press, Cambridge, 2009). \url{https://doi.org/10.1017/CBO9780511626357}

\bibitem{Trefethen2000}
L. N. Trefethen, \textit{Spectral Methods in MATLAB} (SIAM, Philadelphia, 2000). \url{https://doi.org/10.1137/1.9780898719598}

\bibitem{Hairer2003}
E. Hairer, C. Lubich, and G. Wanner, ``Geometric numerical integration illustrated by the St\"{o}rmer-Verlet method,'' Acta Numer. \textbf{12}, 399 (2003). \url{https://doi.org/10.1017/S0962492902000144}

\bibitem{Yoshida1990}
H. Yoshida, ``Construction of higher order symplectic integrators,'' Phys. Lett. A \textbf{150}, 262 (1990). \url{https://doi.org/10.1016/0375-9601(90)90092-3}

\bibitem{Leimkuhler2009}
B. Leimkuhler and S. Reich, \textit{Simulating Hamiltonian Dynamics} (Cambridge University Press, Cambridge, 2009). \url{https://doi.org/10.1017/CBO9780511614118}

\bibitem{Harris2020}
C. R. Harris \textit{et al}., ``Array programming with NumPy,'' Nature \textbf{585}, 357 (2020). \url{https://doi.org/10.1038/s41586-020-2649-2}

\bibitem{Virtanen2020}
P. Virtanen \textit{et al}., ``SciPy 1.0: Fundamental algorithms for scientific computing in Python,'' Nat. Methods \textbf{17}, 261 (2020). \url{https://doi.org/10.1038/s41592-019-0686-2}

\bibitem{Rew1990}
R. Rew and G. Davis, ``NetCDF: An interface for scientific data access,'' IEEE Comput. Graph. Appl. \textbf{10}, 76 (1990). \url{https://doi.org/10.1109/38.56302}

\bibitem{Hunter2007}
J. D. Hunter, ``Matplotlib: A 2D graphics environment,'' Comput. Sci. Eng. \textbf{9}, 90 (2007). \url{https://doi.org/10.1109/MCSE.2007.55}

\bibitem{Lam2015}
S. K. Lam, A. Pitrou, and S. Seibert, ``Numba: A LLVM-based Python JIT compiler,'' in \textit{Proceedings of the Second Workshop on the LLVM Compiler Infrastructure in HPC} (ACM, New York, 2015), pp. 1--6. \url{https://doi.org/10.1145/2833157.2833162}

\bibitem{Herho2024Eks}
S. Herho, S. N. Kaban, D. E. Irawan, and R. Kapid, ``Efficient 1D heat equation solver: Leveraging Numba in Python,'' Eksakta \textbf{25}, 126 (2024). \url{https://doi.org/10.24036/eksakta/vol25-iss02/487}

\bibitem{Herho2025Schrodinger}
S. H. S. Herho, I. P. Anwar, F. Khadami, R. Suwarman, and D. E. Irawan, ``\texttt{simple-idealized-1d-nlse}: Pseudo-spectral solver for the 1D nonlinear Schr\"{o}dinger equation,'' arXiv (2025). \url{https://doi.org/10.48550/arXiv.2509.05901}

\bibitem{Herho2025KH2D}
S. H. S. Herho \textit{et al}., ``\texttt{kh2d-solver}: A Python library for idealized two-dimensional incompressible Kelvin-Helmholtz instability,'' arXiv (2025). \url{https://doi.org/10.48550/arXiv.2509.16080}

\bibitem{Zakharov2001}
V. E. Zakharov, P. Guyenne, A. N. Pushkarev, and F. Dias, ``Wave turbulence in one-dimensional models,'' Physica D \textbf{152--153}, 573 (2001). \url{https://doi.org/10.1016/S0167-2789(01)00194-4}

\bibitem{Sugiyama1979}
T. Sugiyama, ``Kink-antikink collisions in the two-dimensional $\phi^4$ model,'' Prog. Theor. Phys. \textbf{61}, 1550 (1979). \url{https://doi.org/10.1143/PTP.61.1550}

\bibitem{Ram2020}
M. Ram (Ed.), \textit{Recent Advances in Mathematics for Engineering} (CRC Press, Boca Raton, 2020). \url{https://doi.org/10.1201/9780429200304}

\bibitem{Kivshar2002}
Y. S. Kivshar and G. I. Stegeman, ``Spatial optical solitons,'' Opt. Photonics News \textbf{13}, 59 (2002). \url{https://doi.org/10.1364/OPN.13.2.000059}

\bibitem{Fogel1977}
M. B. Fogel, S. E. Trullinger, A. R. Bishop, and J. A. Krumhansl, ``Dynamics of sine-Gordon solitons in the presence of perturbations,'' Phys. Rev. B \textbf{15}, 1578 (1977). \url{https://doi.org/10.1103/PhysRevB.15.1578}

\bibitem{Kivshar1989}
Y. S. Kivshar and B. A. Malomed, ``Dynamics of solitons in nearly integrable systems,'' Rev. Mod. Phys. \textbf{61}, 763 (1989). \url{https://doi.org/10.1103/RevModPhys.61.763}

\bibitem{Gleiser2000}
M. Gleiser and A. Sornborger, ``Long-lived localized field configurations in small lattices: Application to oscillons,'' Phys. Rev. E \textbf{62}, 1368 (2000). \url{https://doi.org/10.1103/PhysRevE.62.1368}

\bibitem{Manton2010}
N. Manton and P. Sutcliffe, \textit{Topological Solitons} (Cambridge University Press, Cambridge, 2010). \url{https://doi.org/10.1017/CBO9780511617034}

\bibitem{Goldstein2002}
H. Goldstein, C. Poole, and J. Safko, \textit{Classical Mechanics}, 3rd ed. (Addison-Wesley, San Francisco, 2002).

\bibitem{Landau1975}
L. D. Landau and E. M. Lifshitz, \textit{The Classical Theory of Fields}, 4th ed. (Butterworth-Heinemann, Oxford, 1975).

\bibitem{Noether1918}
E. Noether, ``Invariante Variationsprobleme,'' Nachr. Ges. Wiss. G\"{o}ttingen Math.-Phys. Kl. \textbf{1918}, 235 (1918).

\bibitem{Shannon1948}
C. E. Shannon, ``A mathematical theory of communication,'' Bell Syst. Tech. J. \textbf{27}, 379 (1948). \url{https://doi.org/10.1002/j.1538-7305.1948.tb01338.x}

\bibitem{Renyi1961}
A. R\'{e}nyi, ``On measures of entropy and information,'' in \textit{Proceedings of the Fourth Berkeley Symposium on Mathematical Statistics and Probability} (University of California Press, Berkeley, 1961), Vol. 1, pp. 547--561.

\bibitem{Tsallis1988}
C. Tsallis, ``Possible generalization of Boltzmann-Gibbs statistics,'' J. Stat. Phys. \textbf{52}, 479 (1988). \url{https://doi.org/10.1007/BF01016429}

\bibitem{Borges2004}
E. P. Borges, ``A possible deformed algebra and calculus inspired in nonextensive thermostatistics,'' Physica A \textbf{340}, 95 (2004). \url{https://doi.org/10.1016/j.physa.2004.03.082}

\bibitem{Silverman1986}
B. W. Silverman, \textit{Density Estimation for Statistics and Data Analysis} (Routledge, New York, 1998). \url{https://doi.org/10.1007/978-1-4899-3324-9}

\bibitem{Kruskal1952}
W. H. Kruskal and W. A. Wallis, ``Use of ranks in one-criterion variance analysis,'' J. Am. Stat. Assoc. \textbf{47}, 583 (1952). \url{https://doi.org/10.1080/01621459.1952.10483441}

\bibitem{Cliff1993}
N. Cliff, ``Dominance statistics: Ordinal analyses to answer ordinal questions,'' Psychol. Bull. \textbf{114}, 494 (1993). \url{https://doi.org/10.1037/0033-2909.114.3.494}

\bibitem{Romano2006}
J. Romano, J. D. Kromrey, J. Coraggio, and J. Skowronek, ``Exploring methods for evaluating group differences on the NSSE and other surveys: Are the $t$-test and Cohen's $d$ indices the most appropriate choices?,'' in \textit{Annual Meeting of the Southern Association for Institutional Research} (Arlington, VA, 2006), pp. 1--33.

\bibitem{Arnold1989}
V. I. Arnold, \textit{Mathematical Methods of Classical Mechanics}, 2nd ed. (Springer-Verlag, New York, 1989). \url{https://doi.org/10.1007/978-1-4757-2063-1}

\end{thebibliography}
\end{document}